# A projection pursuit framework for testing general high-dimensional hypothesis

Yinchu Zhu

Rady School of Management, University of California, San Diego

and

Jelena Bradic *

Department of Mathematics, University of California, San Diego

May 2, 2017

## Abstract

This article develops a framework for testing general hypothesis in high-dimensional models where the number of variables may far exceed the number of observations. Existing literature has considered less than a handful of hypotheses, such as testing individual coordinates of the model parameter. However, the problem of testing general and complex hypotheses remains widely open. We propose a new inference method developed around the hypothesis adaptive projection pursuit framework, which solves the testing problems in the most general case. The proposed inference is centered around a new class of estimators defined as $l_1$ projection of the initial guess of the unknown onto the space defined by the null. This projection automatically takes into account the structure of the null hypothesis and allows us to study formal inference for a number of long-standing problems. For example, we can directly conduct inference on the sparsity level of the model parameters and the minimum signal strength. This is especially significant given the fact that the former is a fundamental condition underlying most of the theoretical development in high-dimensional statistics, while the latter is a key condition used to establish variable selection properties. Moreover, the proposed method is asymptotically exact and has satisfactory power properties for testing very general functionals of the high-dimensional parameters. The simulation studies lend further support to our theoretical claims and additionally show excellent finite-sample size and power properties of the proposed test.

*Keywords:* high-dimensional inference, linear model, generalized linear model, non-convexity, robust test, p-values, bootstrap

---

*The authors gratefully acknowledge the support of NSF DMS award #1205296

# 1 Introduction

High-dimensional statistical inference of models in which the number of variables $p$ might be much larger than the sample size $n$ (i.e., $p \gg n$) has become a fundamental issue in many areas of applications. Examples include image analysis, analysis of high-throughput genomic sequences, speech analysis, etc where the HDLSS (high-dimensional low sample size with $p/n \to \infty$) data structure is apparent. The goal of many empirical analyses is to understand the parameter structure of the model at hand, hence developing methodology that is flexible and broad-ranging as far as the hypotheses are concerned, is of great practical importance. Our goal is to construct a test that can perform well for a large class of high-dimensional null hypothesis.

In this paper, we consider a general model based on $n$ independent and identically distributed observations $z_1, z_2, \ldots, z_n$ of a random variable (vector) $z$ with support $\mathcal{Z}$, a loss function $l(\cdot, \cdot) : \mathcal{Z} \times \mathcal{B} \to \mathbb{R}$ a parameter space $\mathcal{B} \subseteq \mathbb{R}^p$ and a true parameter value defined as

$$\beta_* = \arg\min_{\beta \in \mathcal{B}} L(\beta), \tag{1.1}$$

with $L(\beta) = \mathbb{E}[l(z, \beta)]$. The above formulation covers many important statistical models; for the parametric likelihood models, $z$ is generated from a distribution $P_\beta$ with the true value $\beta_*$ defined by (1.1) with $l(z, \beta) = -\log p_\beta(z)$, where $p_\beta(\cdot)$ is the probability density function corresponding to $P_\beta$.

The main goal of this paper is to fill in the gap in the current high-dimensional literature and present a comprehensive methodology for the following testing problem

$$H_0 : \beta_* \in \mathcal{B}_0 \quad \text{vs} \quad H_1 : \beta_* \notin \mathcal{B}_0, \tag{1.2}$$

for a given set $\mathcal{B}_0 \subset \mathcal{B}$. Here, no restrictions, such as convexity or dimensionality, are imposed on the set $\mathcal{B}_0$. Explicitly, we would like to design a test that is asymptotically exact irrespective of the geometry of the set $\mathcal{B}_0$.

This problem can be motivated first as a high-level approach to performing inference in high-dimensions for complex hypothesis. Since $p \gg n$, further assumptions, such is sparsity have been naturally exploited. There one assumes that the number of non-zeros of $\beta_*$, $\|\beta_*\|_0$, is smaller than $n$. Despite the fact that hypothesis testing problem (1.2) is a difficult problem, it is naturally related to a number of important questions.

**Example 1.1 (Testing the sparsity level).** Over the past decade, sparsity assumption has become the prime measure of regularity. In particular, smooth functions can be viewed as functions with a sparse expansion in an appropriate basis. While the estimation of sparse parameters has been well established across many different models (Meinshausen and Bühlmann, 2006; Van de Geer, 2008; Zhang and Huang, 2008; Bickel et al., 2009; Negahban et al., 2012), this paper, to the best of our knowledge, provides the first testing procedure for the sparsity of the model. Indeed, the sparsity assumption is more often than not poorly justified and has been unverified. As such, our work has direct impact on real applications. In the setup (1.2), we consider $\mathcal{B}_0 = \{\beta \in \mathcal{B} \mid \|\beta\|_0 \leq c\}$ for some pre-specified constant $c \geq 0$.

**Example 1.2 (Testing minimum signal strength).** The three main themes in high-dimensional statistics which have been considered in the past are: prediction of the regression surface, estimation of the parameter vector and variable selection. The last theme concerns with finding a set $\hat{\mathcal{S}}$ such that $\mathbb{P}(\hat{\mathcal{S}} = \mathcal{S}_0)$ is large, where $\mathcal{S}_0 = \mathrm{supp}(\beta_*) := \{j \mid \beta_{*,j} \neq 0\}$ is often referred to as the active set. In order for any regularized estimator to be sign consistent for the active set $\mathcal{S}_0$, we have to require that the non-zero regression coefficients are sufficiently large by imposing a "beta-min" assumption whose asymptotic form reads

$$\min_{j \in \mathcal{S}_0} |\beta_{*,j}| \gg \sqrt{\|\beta_*\|_0 \log(p)/n}.$$

The "beta-min" assumption is restrictive and non-checkable. Furthermore, these conditions are necessary (Fan and Li, 2001; Zhao and Yu, 2006; Meinshausen and Yu, 2009; Fan and Lv, 2011). Hence, for real application it is of paramount importance to verify whether "beta-min" condition holds. To the best of our knowledge, our work is the first that provides a valid test that is asymptotically exact. In the setup (1.2), we consider $\mathcal{B}_0 = \{\beta \in \mathcal{B} \mid \min_{j \in \mathrm{supp}(\beta)} |\beta_j| \geq c\}$ for some prespecified $c > 0$. If $\mathrm{supp}(\beta) = \emptyset$, the convention is $\min_{j \in \mathrm{supp}(\beta)} |\beta_j| = +\infty$ – that is, $\beta = 0$ satisfies the "beta-min" condition.

**Example 1.3 (Testing quadratic forms).** Suppose now that our goal is to construct a two-sided confidence interval (CI) for the signal squared magnitude $\|\beta_*\|_2^2$. In linear models, this is related to inference on the signal-to-noise ratio and the noise level; see (Dicker, 2014; Fan et al., 2012; Sun and Zhang, 2012; Janson et al., 2015; Verzelen and Gassiat, 2016). This is a fundamental statistical problem not only for linear models but also for many nonlinear models that are popular in applications. However, this problem has not been discussed much beyond high-dimensional linear models. Our method allows us to test the more general hypotheses of the form

$$H_0: \|Q\beta_*\|_2 \in A$$

for some pre-specified set $A \subseteq [0, \infty)$ and a matrix $Q \in \mathbb{R}^{p \times p}$, where $\beta_*$ is the parameter in a general model defined in (1.1). In the setup (1.2), we shall consider $\mathcal{B}_0 = \{\beta \in \mathcal{B} \mid \|Q\beta\|_2 \leq c\}$ for some prespecified $c > 0$.

Many important problems in practice reduce to the inference problem of complex hypotheses in nonlinear models. Consider the problem of testing $H_0 : \psi(\beta_*) \leq 0$ for some functional $\psi(\cdot)$ defined on $\mathcal{B}$. In our framework, this hypothesis is equivalent to (1.2) with $\mathcal{B}_0 = \{\beta \in \mathcal{B} \mid \psi(\beta) \leq 0\}$. The literature has seen some progress in this direction for linear models, such as Cai and Guo (2015) for linear functionals, however, more general functionals have not been discussed yet. A practical example is whether an individual with feature $x_1$ is twice as likely as the one with feature $x_2$ to contract the disease (event). In this case the functional of interest would be $\psi(\beta) = g(x_1^\top \beta) - 2g(x_2^\top \beta)$ in the logistic regression model where $g(\cdot)$ is the distribution function for the standard logistic distribution. Notice that when $\mathcal{B}_0$ is written in terms of a functional on $\beta_*$, we do not require properties such as convexity or smoothness in the functional under testing.

In this article, we propose a method that can be used for general hypotheses (1.2), including all the aforementioned inference problems. We start with the testing problem (1.2) in linear models and extend the methodology to various regression models. The test we propose uses bootstrap to obtain critical values and is computationally simple as we do not need to re-estimate the model in bootstrap samples. Under mild regularity conditions, we show that our test provides an asymptotically exact inference procedure and also possesses satisfactory power properties.

## 1.1 Related Work

A significant understanding has emerged over the past few years that statistical inference of parameters with dimension much larger than the sample size can be problematic in the situation of an "imperfect model selection", which can arise when the features corresponding to the true model parameters and the nuisance parameters have a high degree of correlation (Zhao and Yu, 2006). The situation of parameter estimation has been studied extensively in the presence of "imperfect model selection" (Bunea et al., 2007; Bickel et al., 2009); however, the literature on high-dimensional testing of complex hypothesis is very light. In the context of high dimensional linear models, Zhang and Zhang (2014) have introduced tests to compare a single variable to a prescribed value. In the spirit of the Wald test, their methodology named "debiasing" is based on exploiting optimal low-dimensional projections of the parameter of the model. The methodology has been extended to problem-specific settings of generalized linear models (Van de Geer et al., 2014), gaussian graphical models (Ren et al., 2013), matrix estimation (Cai et al., 2014; Janková and van de Geer, 2015) and etc. Ning and Liu (2014) developed a parallel approach, named score tests, based on optimal projections of the score vector instead. A competing method of Javanmard and Montanari (2014) guarantees a test of an asymptotically exact size and relaxes the sparsity of the inverse covariance matrix. The authors provide an optimal sample size computation for their method in Javanmard and Montanari (2015). Recently, Zhang and Cheng (2016) and Dezeure et al. (2016), adopted a gaussian multiplier bootstrap approach of Chernozhukov et al. (2013) in an adaptation of the debiasing procedure to the simultaneous testing of groups of variables. However, existing literature only considers only simple hypotheses that specify one or some of the entries of the parameter to be given values. In contrast, complex null hypotheses (both convex and non-convex) – hypotheses that depend on all the entries of $\beta_*$, not just a few and those that allow general interactions between all elements of $\beta_*$ – in high-dimensional models have presented significant challenges and have not been successfully solved. Moreover, we observe that these existing approaches do not extend naively to general complex null hypotheses, due to error accumulation in high-dimensions; thus, new methodology is required. Meinshausen (2015) and Mandozzi and Bühlmann (2016) build the testing strategy upon the knowledge of the distributional form of the error term and design tests adaptive to highly correlated designs. However, their approach is not asymptotically exact and provides conservative inference bounds.

In addition to being able to test general functionals of $\beta_*$, what methodologically distinguishes our treatment from the above existing literature, is that our proposal is not centered around a construction of an unbiased estimator or unbiased score equations, as those would not be possible for many functionals of interest (including but not limited to the Examples 1.1–1.3) – estimation of non-smooth functionals is particularly hard even for low-dimensional problems. In fact, it is not clear what one should estimate to test the general hypothesis (1.2). To overcome the inherited difficulty of estimation, we create a projection pursuit suited for testing problems, by infusing the null hypothesis set $\mathcal{B}_0$ into the construction of the $\ell_1$-projection of a suitable estimator of $\beta_*$. When $\mathcal{B}_0$ is defined in terms of a functional $\psi(\cdot)$ of $\beta_*$, our methodology allows us to bypass the estimation of $\psi(\beta_*)$ and therefore relax many assumptions on the functional $\psi(\cdot)$. Further details involve then a construction of a test statistic related to the size of the residuals of the projection pursuit procedure.

## 1.2 Notations and organization of the paper

Throughout this paper, $^\top$ denotes the matrix transpose and $\mathbb{I}_p$ denotes the $p \times p$ identity matrix with its $j$th column denoted by $e_j$. The (multivariate) Gaussian distribution with mean (vector) $\mu$ and variance (matrix) $\Sigma$ is denoted by $N(\mu, \Sigma)$. For a vector $v \in \mathbb{R}^k$, we denote its $j$th entry by $v_j$ and define its $\ell_q$-norm as follows: $\|v\|_q = (\sum_{i=1}^k |v_i|^q)^{1/q}$ for $q \in (0, \infty)$, $\|v\|_\infty = \max_{1 \leq i \leq k} |v_i|$ and $\|v\|_0 = \sum_{i=1}^k \mathbf{1}\{v_i = 0\}$, where $\mathbf{1}\{\}$ denotes the indicator function. We also define $\mathrm{supp}(v) = \{j \mid |v|_j > 0\}$. For two sequences $a_n, b_n > 0$, we use $a_n \asymp b_n$ to denote that there exist positive constants $C_1, C_2 > 0$ such that $\forall n$, $a_n \leq C_1 b_n$ and $b_n \leq C_2 a_n$. For two real numbers $a$ and $b$, let $a \vee b$ and $a \wedge b$ denote $\max\{a, b\}$ and $\min\{a, b\}$, respectively. For a differentiable real-valued function $f(\cdot)$ of $x$, we define $\nabla_x f(x) = \partial f(x)/\partial x$ and $\nabla_x^2 f(x) = \partial^2 f(x)/\partial x \partial x^\top$. We use "s.t." as the abbreviation for "subject to".

We also introduce two definitions that will be used frequently. The sub-Gaussian norm of a random variable $X$ is defined as $\|X\|_{\psi_2} = \sup_{q \geq 1} q^{-1/2} (\mathbb{E}|X|^q)^{1/q}$, whereas the sub-Gaussian norm of a random vector $Y \in \mathbb{R}^k$ is $\|Y\|_{\psi_2} = \sup_{\|v\|_2 = 1} \|v^\top Y\|_{\psi_2}$. The sub-exponential norm of a random variable $X$ is defined as $\|X\|_{\psi_1} = \sup_{q \geq 1} q^{-1} (\mathbb{E}|X|^q)^{1/q}$. A random variable is said to be sub-Gaussian (sub-exponential) if its sub-Gaussian (sub-exponential) norm is finite.

The rest of the paper is structured as follows. Section 2 introduces $\ell_1$ Projection Pursuit estimator, its properties and guidelines on the implementation. Section 3 presents the Projection Pursuit Testing methodology for the linear model and its theoretical properties. General Projection Pursuit Testing methodology is introduced in Section 4 as well as its theoretical properties. In Section 5, we illustrate the proposed methodology on the class of generalized linear models through a case of the logistic model. In Section 6, we demonstrate the finite-sample performance in Monte Carlo simulations. The proofs for the theoretical results are contained in the Supplementary Materials.

## 2 $\ell_1$-projection pursuit estimator

In this section we introduce a new class of estimators, named $\ell_1$-*projection pursuit estimators* designed as hypothesis adaptive contrasts for the testing problem (1.2). Projection pursuit ideas originally developed for the multivariate analysis of high-dimensional point clouds, were designed to "pick" interesting low-dimensional projections of a high-dimensional observations; see Friedman and Tukey (1974) and Huber (1985) for example. Here, we propose a class of estimators in high-dimensional setting useful for detecting deviations from the null hypothesis – they are designed to "pick" useful projections of the unknown parameter to the null hypothesis set $\mathcal{B}_0$ that is of our interest.

We measure the deviations from $H_0$ using the Hausdorff distance (based on the $\ell_1$-norm), i.e., $d(\mathcal{B}_0, \beta_*) = \min_{v \in \mathcal{B}_0} \|\beta_* - v\|_1$. Observe that the null $H_0 : \beta_* \in \mathcal{B}_0$ is equivalent to the hypothesis that $d(\mathcal{B}_0, \beta_*) = 0$. As $\beta_*$ is unknown, direct computation of this distance is not feasible. Instead, we replace the unknown $\beta_*$ with its candidate estimate $\hat{\beta}_u$ developed by utilizing observations $z_1, \ldots, z_n$.

This estimator is hypothesis blind, i.e. it is unconstrained by the definition of the null; hence, the notation $\hat{\beta}_u$. For example, in linear models, we can take as $\hat{\beta}_u$ the famous Dantzig selector, defined as the solution to the following optimization problem

$$\hat{\beta}_u = \arg\min_{\beta \in \mathbb{R}^p} \left\{ \|\beta\|_1 \ s.t. \ \|n^{-1}X^\top(Y - X\beta)\|_\infty \leq \lambda \right\}, \tag{2.1}$$

where $\lambda \asymp \sqrt{n^{-1} \log p}$ is a tuning parameter specifying the amount of sparsity to be encouraged at estimation.

The element in $\mathcal{B}_0$ that matches the Hausdorff distance $d(\mathcal{B}_0, \hat{\beta}_u)$ is referred to as the *projection pursuit estimator*, i.e., the solution to the following optimization problem

$$\begin{aligned} \hat{\beta}_d = \arg\min_{\beta \in \mathbb{R}^p} & \quad \|\beta - \hat{\beta}_u\|_1 \\ \text{s.t.} & \quad \beta \in \mathcal{B}_0. \end{aligned} \tag{2.2}$$

The distance $\|\hat{\beta}_d - \hat{\beta}_u\|_1$ serves as an estimate for $d(\mathcal{B}_0, \beta_*)$, which takes value zero under $H_0$. Such a distance measure is a random variable, whose distribution can provide a benchmark to gauge whether the deviations of the null hypothesis $\beta_* \in \mathcal{B}_0$ are any better than a random chance. The choice of $\ell_1$-norm in the Hausdorff distance is motivated by the fact that for high-dimensional problems with $p \gg n$, the $\ell_1$-norm typically induces sparsity and enables consistent estimation. The following result says that for an $\ell_1$-consistent estimator $\hat{\beta}_u$, $\|\hat{\beta}_u - \beta_*\|_1 = 0_P(1)$, the introduced $\hat{\beta}_d$ is also $\ell_1$-consistent whenever the null hypothesis is true.

**Lemma 1.** *If the null hypothesis $H_0 : \beta_* \in \mathcal{B}_0$ holds, the $\ell_1$ projection pursuit estimator $\hat{\beta}_d$ (2.2) satisfies $\|\hat{\beta}_d - \beta_*\|_1 \leq 2\|\hat{\beta}_u - \beta_*\|_1$.*

The $\ell_1$-norm is special in the sense that the above consistency property does not hold if we replace the $\ell_1$-norm in (2.2) with $\ell_q$-norms for any $q \neq 1$. Below we illustrate the geometry of the proposed $\ell_1$ projection pursuit estimator for complex null sets $\mathcal{B}_0$, via examples introduced in Section 1. We observe that for many non-convex sets $\mathcal{B}_0$ the proposed estimator possesses analytical solution – hence, bypassing the difficult non-convex optimization.

## 2.1 Example 1.1 (continued)

Although the $\ell_0$-ball $\mathcal{B}_0 = \{\beta \mid \|\beta\|_0 \leq s_0\}$ is not convex, the computational burden is in fact negligible due to exploitation of the geometric structure of $\mathcal{B}_0$. The following result characterizes projection onto "$\ell_0$-balls" as the thresholding rule of keeping the largest entries and setting the others to zero.

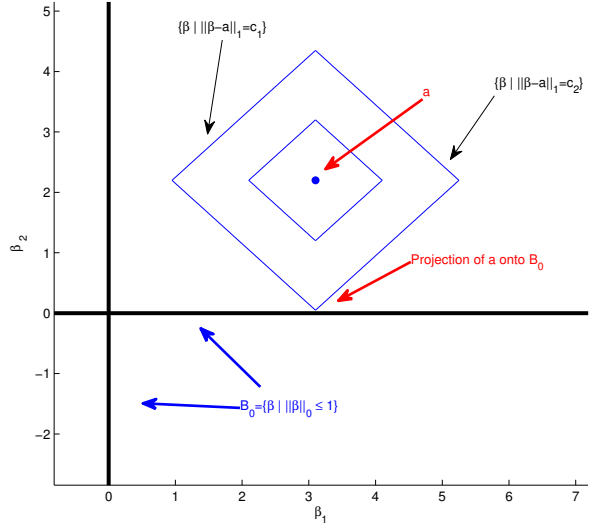

Figure 2.1: $\ell_1$-projection onto $\ell_0$-balls in $\mathbb{R}^2$ where the black lines represent the $\ell_0$-ball $\mathcal{B}_0 = \{\beta \mid \|\beta\|_0 \leq 1\}$ and rhombi centered at $a$ denote the level sets $\{\beta \mid \|\beta - a\|_1 = c\}$.

**Lemma 2.** *Let $v \in \mathbb{R}^p$ and $s_0$ be a nonnegative integer. Suppose that $\pi : \{1, \cdots, p\} \to \{1, \cdots, p\}$ is a permutation such that $|v_{\pi(1)}| \geq |v_{\pi(2)}| \geq \cdots \geq |v_{\pi(p)}|$. Define $\tilde{v} \in \mathbb{R}^p$ with $\tilde{v}_J = v_J$ and $\tilde{v}_{J^c} = 0$ with $J = \{\pi(1), \cdots, \pi(s_0)\}$. Then $\tilde{v}$ solves $\min_{\beta \in \mathbb{R}^p} \|\beta - v\|_1$ s.t. $\|\beta\|_0 \leq s_0$.*

By Lemma 2, the computation of $\ell_1$-projection pursuit estimator is extremely simple and does not require any numerical optimization. Figure 2.1 illustrates the geometry of the computation for $p = 2$ and $s_0 = 1$. To appreciate the convenience of the closed-end form solution in Lemma 2, we note that optimization over $\ell_0$-balls are in general very challenging. For example, if we replace the $\ell_1$-norm with $\ell_2$-norm in (2.2) we would lose this property as the high-dimensional linear models can be estimated using an $\ell_0$-penalized least squared method, for which the computation is known to be NP-hard.

## 2.2 Example 1.2 (continued)

We begin by observing that the beta-min set $\mathcal{B}_0 = \{\beta \in \mathcal{B} \mid \min_{j \in \text{supp}(\beta)} |\beta_j| \geq c\}$ is highly nonconvex, as illustrated in Figure 2.2 (left) with $p = 2$. Fortunately, the computation can be done efficiently due to the following result, which provides a closed-end solution to the optimization problem.

**Lemma 3.** *Let $v \in \mathbb{R}^p$ and $c > 0$. Define $\tilde{v} \in \mathbb{R}^p$ with $\tilde{v}_j = \rho(v_j, c) \; \forall 1 \leq j \leq p$, where $\rho(a, c) = \text{sign}(a)\mathbf{1}\{c/2 \leq |a| < c\}c + \mathbf{1}\{|a| \geq c\}a$. Then $\tilde{v}$ solves $\min_{\beta \in \mathbb{R}^p} \|\beta - v\|_1$ s.t. $\min_{j \in \text{supp}(\beta)} |\beta_j| \geq c$.*

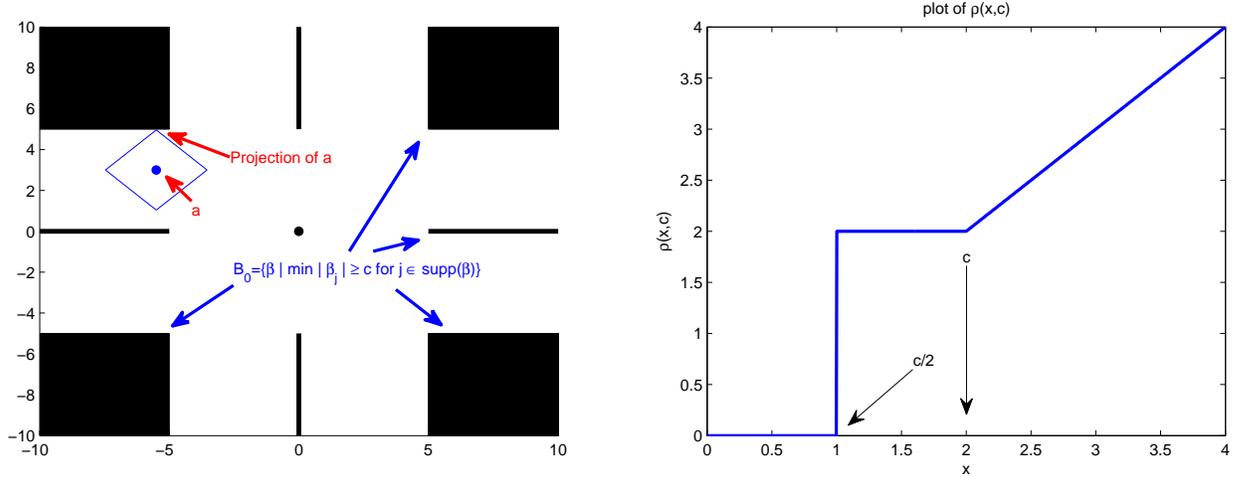

Figure 2.2: $\ell_1$-projection onto the set $\mathcal{B}_0 = \{\beta \in \mathcal{B} \mid \min_{j\in\text{supp}(\beta)} |\beta_j| \geq c\}$ (left) and the thresholding function $\rho(\cdot, c)$ (right).

The function $\rho(\cdot, c)$ can be viewed as a thresholding rule. Entries that are larger than $c$ in magnitude are unchanged; otherwise, entries smaller than $c/2$ in magnitude are set to zero and those with magnitude between $c/2$ and $c$ are set to $c$ or $-c$, depending on the sign (see Figure 2.2 (right)). Lemma 3 says that minimizing $\ell_1$-distance over the beta-min set amounts to applying the thresholding rule entrywise.

## 2.3 Example 1.3 (continued)

Figure 2.3 illustrates $l_1$ projection onto the set $\mathcal{B}_0 = \{\beta \in \mathcal{B} \mid \|Q\beta\|_2 \leq r_0\}$ with $p = 2$ and $Q = \mathbb{I}_p$. Notice that this is a convex optimization problem and thus it can be solved by efficient algorithms. The result below provides an easy method that exploits fast computation packages that are widely available for statistical analysis. Applying the duality theory, we show that the optimization problem can be reduced to a one-dimensional problem once the solution path of a certain Lasso problem is obtained.

**Lemma 4** (Example 1.3 continued). *Define the function $t \mapsto a(t)$ with $a(t) \in \arg\min_{a\in\mathbb{R}^p} \|Qv + Qa\|_2^2 + t\|a\|_1$. Let $t_* \geq 0$ solve $\min_{t\geq 0} \|a(t)\|_1$ subject to $\|Qv + Qa(t)\|_2 \leq c$. Then $v + a(t_*)$ solves $\min_{\beta \in \mathbb{R}^p} \|\beta - v\|_1$ subject to $\|Q\beta\|_2 \leq c$.*

Notice that the function $a(\cdot)$ in Lemma 4 is the solution path of the Lasso regression in which $Qv$ is the response vector and $-Q$ is the design matrix. Very efficient algorithms have been developed, such as LARS by Efron et al. (2004). Once this solution path is computed Lemma 4 says that $\ell_1$-projection pursuit estimation over an ellipsoid amounts to solving an optimization problem in $t$, a scalar. We illustrate this projection on Figure 2.3.

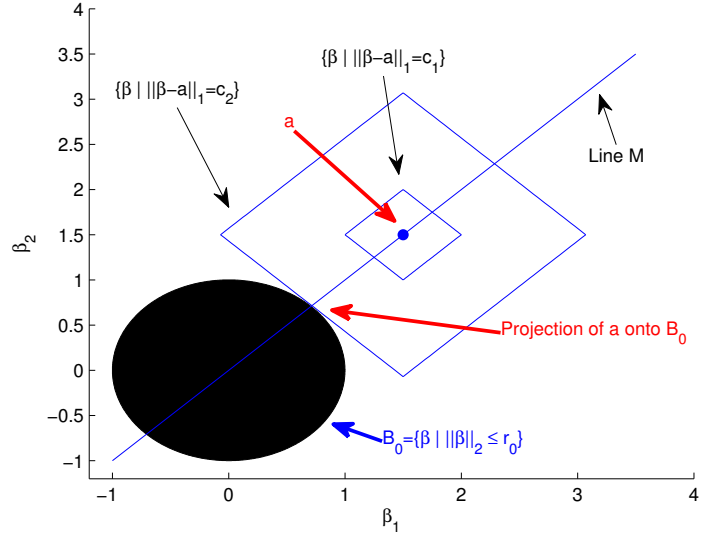

Figure 2.3: $\ell_1$-projection of $a$ onto $\ell_2$-balls $\mathcal{B}_0 = \{\beta \mid \|\beta\|_2 \leq r_0\} \subset \mathbb{R}^2$.

## 3 Projection pursuit test statistic: the case of linear model

Although we will ultimately argue that our method applies much more broadly, we will begin the exposition of the method by considering the following linear model,

$$Y = X\beta_* + \varepsilon, \tag{3.1}$$

where $Y = (y_1, \cdots, y_n)^\top \in \mathbb{R}^n$ is the response and the error $\varepsilon = (\varepsilon_1, \cdots, \varepsilon_n)^\top \in \mathbb{R}^n$ is independent of the design matrix $X = (x_1, \cdots, x_n)^\top \in \mathbb{R}^{n \times p}$ with independent rows $x_i \in \mathbb{R}^p$ and such that $\mathbb{E}[x_i] = 0$, $\mathbb{E}[\varepsilon] = 0$ and $\mathbb{E}[\varepsilon\varepsilon^\top] = \mathbb{I}_n \sigma_\varepsilon^2$. We consider a random design setting, allow for $p \gg n$ and do not assume that the error variance $\sigma_\varepsilon$ is known. Our method will involve data splitting and we use $X_A = (x_1, \cdots, x_m)^\top \in \mathbb{R}^{m \times p}$ and $X_B = (x_{m+1}, \cdots, x_n)^\top \in \mathbb{R}^{m \times p}$ with $m = n/2$ to denote the two subsamples considered later on.

In this section we develop a novel method for constructing test statistics for testing the hypothesis

$$H_0 : \beta_* \in \mathcal{B}_0, \text{ vs. } H_1 : \beta_* \notin \mathcal{B}_0.$$

Our framework does not require $\mathcal{B}_0$ to be a convex set. The only restriction we impose is that it is a closed set; hence, it's size may grow with $p$, its contours may be irregular, it does not have to be a convex set, etc. The methodology we develop here is designed to test the implications of a null hypothesis, whereas the Wald, Score or Likelihood ratio tests, aim to exploit specific parametric restrictions. The score tests, for example, "eliminate" the set of nuisance parameters, whereas our procedure "eliminates" the set of parameters of primary interest. We achieve that by designing a *projection pursuit estimator* that directly utilizes the set $\mathcal{B}_0$ in the construction of the projection, hence avoiding a consistent estimation of the possibly complicated functional of $\beta_*$ describing the set $\mathcal{B}_0$.

## 3.1 Construction of the test statistic

We now wish to introduce the test statistic $T_n$, which will help us contrast the two estimators, the initial one $\hat{\beta}_u$ and the newly constructed and hypothesis dependent, $\ell_1$-projection pursuit estimate $\hat{\beta}_d$. The test $T_n$ is constructed so that large positive values are evidence against the null hypothesis $\beta_* \in \mathcal{B}_0$. One possible candidate would be a test defined as $\|\hat{\beta}_u - \hat{\beta}_d\|_\infty$. However, for $p \gg n$, this is not a good choice since the regularization in estimation can induce asymptotic degeneracy, making the test impossible. Instead, we propose to consider the following test statistic, the *projection pursuit test* (PPTest) statistic, defined as

$$T_n = \sqrt{n} \max_{1 \leq j \leq p} \left| \hat{\beta}_{u,j} - \hat{\beta}_{d,j} - \hat{\delta}_j \right|, \qquad (3.2)$$

where $\hat{\delta} \in \mathbb{R}^p$ is a data-dependent vector, a suitably chosen estimate of the bias of the projection residuals, i.e. such that under the null hypothesis $H_0$, $\hat{\beta}_u - \hat{\beta}_d - \hat{\delta}$ "approximately" has zero mean.

Under the general methodology that we propose, possibly other measures of the size of $\hat{\beta}_u - \hat{\beta}_d - \hat{\delta}$ can be allowed. The $\ell_\infty$-norm allows for extremely large number of model parameters $p$, i.e., $p \gg n$; for instance, instead of considering the $\ell_\infty$-norm, we can evaluate the $\ell_2$-norm whenever $p = o(n)$, that is we can consider a test based on $\left\{ n \sum_{j=1}^p \left( \hat{\beta}_{u,j} - \hat{\beta}_{d,j} - \hat{\delta}_j \right)^2 \right\}$.

At the first glance, one might think that the construction of $\hat{\delta}$ depends on the model, the estimation procedure of $\hat{\beta}_u$ and the set $\mathcal{B}_0$. However, we advocate for a a simple unified method that only takes into an account the model structure. Namely, we define the estimator of the projection bias of the residuals, as the following vector $\hat{\delta} = \hat{\delta}_{SP} \mathbf{1} \left\{ \|\hat{\delta}_{SP}\|_\infty \leq n^{-1/4} \right\}$ with

$$\hat{\delta}_{SP} = \hat{\Theta} m^{-1} \sum_{i=1}^m x_i(y_i - x_i^\top \hat{\beta}_u) - \hat{\Theta} m^{-1} \sum_{i=m+1}^n x_i(y_i - x_i^\top \hat{\beta}_d) \qquad (3.3)$$

where the matrix $\hat{\Theta} \in \mathbb{R}^{p \times p}$ is defined as $\hat{\Theta} = (\hat{\Theta}_1, \cdots, \hat{\Theta}_p)^\top \in \mathbb{R}^{p \times p}$. Here, we keep the smallest few elements of $\hat{\delta}_{SP}$ to improve the sensitivity of the proposed test. Moreover, the thresholding level that we advocate is universal, $n^{-1/4}$ and does not need to be tuned in finite samples; i.e., the method is not sensitive to big changes. Finally, for $1 \leq j \leq p$, the columns $\hat{\Theta}_j$ are the solutions to the following problem

$$\hat{\Theta}_j = \underset{\theta \in \mathbb{R}^p}{\arg\min} \|\theta\|_1 \quad s.t. \quad \|m^{-1} X_A^\top X_A \theta - e_j\|_\infty \leq \eta \qquad (3.4)$$
$$\|m^{-1} X_B^\top X_B \theta - e_j\|_\infty \leq \eta$$
$$\|X\theta\|_\infty \leq \mu.$$

In the above definition, tuning parameters are chosen as $\eta \asymp \sqrt{n^{-1} \log p}$ and $\mu \asymp \sqrt{\log(p \vee n)}$. Observe that $\hat{\Theta}$ is constructed as a pooled variance estimator modifying the CLIME estimator introduced by Cai et al. (2011). Alternatively, we can follow a similar algorithm as in Javanmard and Montanari (2014). The node-wise Lasso studied by Meinshausen and Bühlmann (2006) can also be used if we assume sparsity in rows of $(\mathbb{E} x_i x_i^\top)^{-1}$.

## 3.2 Construction of the critical value

Due to highly complicated dependencies in high-dimensions, obtaining an asymptotic pivotal distribution of $T_n$ is difficult, if at all possible. Moreover, observe that the introduced test statistic $T_n$ changes its form with the change of the null set $\mathcal{B}_0$ via the introduced $\ell_1$-projection pursuit estimator $\hat{\beta}_d$, therefore introducing additional difficulties in computing its exact distribution. Hence, we instead aim at finding a data-driven critical value of the test statistic $T_n$, i.e. value $c$ such that whenever the null hypothesis $H_0 : \beta_* \in \mathcal{B}_0$ holds, $T_n > c$ holds with probability close to the prescribed level of the test $\alpha \in (0, 1)$.

We introduce an efficient multiplier bootstrap procedure to achieve this goal. Namely, consider the data-driven vectors $\hat{R}_1, \ldots, \hat{R}_n \in \mathbb{R}^p$

$$\widehat{R}_i = \begin{cases} -2\hat{\Theta} x_i (y_i - x_i^\top \hat{\beta}_u) & \text{for } 1 \leq i \leq m \\ 2\hat{\Theta} x_i (y_i - x_i^\top \hat{\beta}_u) & \text{for } m+1 \leq i \leq n \end{cases}, \qquad (3.5)$$

together with their empirical average $R^* = n^{-1} \sum_{i=1}^n \widehat{R}_i \in \mathbb{R}^p$. Together with them we consider a sequence of independent and identically distributed (i.i.d.) random vectors, the Gaussian multipli-ers $\xi_1, \ldots, \xi_n$, each with a standard normal distribution. We then propose to consider the following bootstrap test statistics

$$T_n^{BS} = n^{-1/2} \max_{1 \leq j \leq p} \left| \sum_{i=1}^n \left( \widehat{R}_{i,j} - R_j^* \right) \xi_i \right|.$$

We will show that the resampled distribution of $T_n^{BS}$ offers a good approximation of the distribution of $T_n$. Hence, we can take as the critical value for $T_n$ the $(1-\alpha)$-quantile of $T_n^{BS}$ (while holding the observations fixed), denoted by $\mathcal{Q}(1-\alpha, T_n^{BS})$, where $\alpha$ is the nominal size of the test.

## 3.3 Theoretical properties

In this subsection we detail theoretical properties of the proposed test.

**Theorem 1.** *For the linear regression model (3.1), consider a test that rejects the null iff $T_n > \mathcal{Q}(1-\alpha, T_n^{BS})$. Suppose that $\log p = o(n^{1/8})$ and $\|\beta_*\|_0 = o(n^{1/4}/\sqrt{\log p})$. Let $\sigma_\varepsilon$ and all the eigenvalues of $\Sigma_X = \mathbb{E}\left[x_1 x_1^\top\right]$ lie in $[\kappa_1, \kappa_2]$ for $\kappa_1, \kappa_2 \in (0, \infty)$. Additionally, assume there exists a constant $\kappa_3 \in (0, \infty)$ such that $\|\Sigma_X^{-1/2} x_1\|_{\psi_2} \leq \kappa_3$ and $\|\varepsilon_1\|_{\psi_2} \leq \kappa_3$. Then, we have that under $H_0$,*

$$\limsup_{n \to \infty} \sup_{\alpha \in (0,1)} \left| \mathbb{P}\left(T_n > \mathcal{Q}(1-\alpha, T_n^{BS})\right) - \alpha \right| = 0.$$

Note that Theorem 1 showcases that the proposed Projection Pursuit Test is an asymptotically $\alpha$ level test, i.e., $P(\text{Type I error}) \to \alpha$. Observe that it allows for high-dimensionality, $p \gg n$, imposes very mild assumptions on design, $\Sigma_X$ and the model error $\varepsilon$. In particular, we do not require $\Sigma_X^{-1}$ to be sparse – a condition utilized in Belloni et al. (2014); Zhang and Cheng (2016) or Belloni et al. (2016). In this way our test proposed a more robust alternative to the simultaneous tests above. To achieve robustness we impose slightly stronger assumptions on sparsity and dimensionality; see the assumption of $\|\beta_*\|_0 = o(\sqrt{n}/\log p)$ in Van de Geer et al. (2014) for example and $\log p = o(n^2)$ of Ning and Liu (2014). The polynomial grow factors are a small price to pay for the complexity of the null hypothesis – observe that the diameter of $\mathcal{B}_0$ can explode with $n$. This is somewhat expected as our null hypothesis for example includes problems of high-dimensional simultaneous testing, as those of Zhang and Cheng (2016); Dezeure et al. (2016); in comparison to those, our restriction on the growth of $p$ is related. We will showcase this general algorithm on all three Examples of Section 1, i.e. testing for sparsity, testing for minimum signal strength and testing for quadratic functionals (see Section 6).

Next, we consider power properties of the proposed test. For that end, we consider the following

alternative

$$H_{1,n}: \min_{\beta \in \mathcal{B}_0} \|\beta - \beta_*\|_\infty \geq c_n,$$

for a suitably chosen positive sequence $c_n$.

**Theorem 2.** *Let the assumptions in Theorem 1 hold. Suppose that $c_n \gg n^{-1/4}$. Then, under $H_{1,n}$, we have that $\forall \alpha \in (0,1)$, $\liminf_{n \to \infty} \mathbb{P}\left(T_n > \mathcal{Q}(1 - \alpha, T_n^{BS})\right) = 1$.*

Theorem 2 says that our test has power against alternatives with deviation larger than $n^{-1/4}$. Hence, our procedure is very sensitive in detecting sparse alternatives. Notice that this rate holds for any $\mathcal{B}_0$ with unknown $\Sigma_X$ and $\sigma_\varepsilon$. Deriving the optimal rate of detection of $H_{1,n}$ is quite difficult even if we take into account the structure of $\mathcal{B}_0$. For example, inference on the $\ell_2$-norm of $\beta_*$ has been studied by Ingster et al. (2010) who derive the minimax detection rate under Gaussian design with $\Sigma_X = I_p$ and unknown $\sigma_\varepsilon$, see Theorem 4.5 and Proposition 4.6 therein; however, to the best of our knowledge, the optimal rate with unknown $\Sigma_X$ for this problem is still an open question. Another example is inference of $H_0: a^\top \beta_* = 0$ with a known dense vector $a$. Cai and Guo (2015) study this problem and derive the minimax rate for tests that use explicit knowledge of $\|\beta_*\|_0$; they also show that tests without explicit knowledge of $\|\beta_*\|_0$ are strictly less powerful in detection rate. The minimax detection rate for this problem without knowledge of the sparsity level is also an open question; in fact, we cannot find such a test that does not use the sparsity level in the literature. Moreover, our test is asymptotically exact and hence might deliver superior performance in finite sample than some minimax tests that are only rate-optimal.

## 4 A Projection Pursuit Methodology

Here we provide a general projection pursuit methodology developed for testing the null $\beta_* \in \mathcal{B}_0$; a methodology that does not require the set $\mathcal{B}_0$ to be convex – with examples extending beyond non-smooth functionals of $\beta_*$, quadratic functionals of $\beta_*$ and many more.

### 4.1 Construction of the test statistic

**Step 1: Compute a consistent estimator of $\beta_*$.** For the observations $z_1, \ldots, z_n$ compute the preliminary estimate, $\hat{\beta}_u$ by minimizing the sample analog of (1.1). We do not impose hard restrictions on how such an estimator is computed, as long as $\|\hat{\beta}_u - \beta_*\|_1 = o_P(1)$. The liberty of choosing from the most convenient estimator is one of the advantages of our method. For high-dimensional problems, we

typically obtain $\hat{\beta}_u$ through a regularized estimation, for example: $\hat{\beta}_u = \arg\min_{\beta \in \mathcal{B}} \{L_n(\beta) + P(\beta)\}$, where $L_n(\cdot)$ is the sample analog of $L(\cdot)$ and $P(\cdot)$ is a penalty function, such as $\lambda \|\cdot\|_1$ in Lasso or the non-convex function in SCAD. One can also use the Dantzig selector.

**Step 2: Construct the $\ell_1$-projection pursuit estimator.** Compute the "projection" of the initial estimate $\hat{\beta}_u$ onto the null set $\mathcal{B}_0$, i.e., $\hat{\beta}_d$ as in (2.2). Observe that this contrast estimate does not depend on the regression model.

**Step 3: Form the test statistic.** Define the *projection pursuit test* (PPTest) statistic as

$$T_n = \sqrt{n} \max_{1 \leq j \leq p} \left| \hat{\beta}_{u,j} - \hat{\beta}_{d,j} - \hat{\delta}_j \right|, \tag{4.1}$$

where for $m = n/2$, $\hat{\delta} = (\hat{\delta}_1, ..., \hat{\delta}_p)^\top = \delta_{SP} \mathbf{1}\{\|\delta_{SP}\|_\infty \leq n^{-1/4}\}$ with

$$\delta_{SP} = m^{-1} \sum_{i=1}^{m} \hat{\Theta}_A s(z_i, \hat{\beta}_u) - m^{-1} \sum_{i=m+1}^{n} \hat{\Theta}_B s(z_i, \hat{\beta}_d), \tag{4.2}$$

In the above display $s(z, \beta) = \nabla_\beta l(z, \beta)$ is the gradient of the loss and $l(\cdot, \cdot)$ is the loss function in (1.1). Matrices $\hat{\Theta}_A, \hat{\Theta}_B \in \mathbb{R}^{p \times p}$ are candidate estimates for the precision matrix $[\nabla^2_\beta \mathbb{E} l(z, \beta_*)]^{-1}$. We note that the particular choice of the suitable estimates $\hat{\Theta}_A$ and $\hat{\Theta}_B$ is problem specific; however, techniques advocated by Zhang and Zhang (2014), Van de Geer et al. (2014) and Javanmard and Montanari (2014) can be adapted to many models. In Section 5, we discuss the particular case of generalized linear models through a logistic model. Moreover, estimates $\hat{\Theta}_A$ and $\hat{\Theta}_B$ can be either pooled ($\hat{\Theta}_A = \hat{\Theta}_B$) or unpooled ($\hat{\Theta}_A \neq \hat{\Theta}_B$); we showcase advantageous properties of both cases in Section 5.

**Step 4: Calculate the critical value via bootstrap.**

Under certain regularity conditions, we show that $T_n$ can be approximated by $n^{-1/2} \|\sum_{i=1}^n R_i\|_\infty$, whenever the null hypothesis holds. Here, the vectors $R_i \in \mathbb{R}^p$ are defined as

$$R_i = \begin{cases} 2\Theta_A s(z_i, \beta_*) & 1 \leq i \leq m \\ -2\Theta_B s(z_i, \beta_*) & m+1 \leq i \leq n. \end{cases} \tag{4.3}$$

We propose to adapt a multiplier bootstrap procedure developed in a series of papers Chernozhukov et al. (2014); Zhang and Cheng (2016); Dezeure et al. (2016). By repeatedly drawing the i.i.d. standard

Gaussian multipliers $\{\xi_i\}_{i=1}^n$, we define the bootstrapped test statistic

$$T_n^{BS} = n^{-1/2} \max_{1 \leq j \leq p} \left| \sum_{i=1}^n \left( \widehat{R}_{i,j} - R_j^* \right) \xi_i \right| \quad \text{with } R_j^* = n^{-1} \sum_{i=1}^n \widehat{R}_{i,j} \quad (4.4)$$

where $\hat{R}_i$ is an suitable estimator for $R_i$. For this purpose, we consider

$$\widehat{R}_i = \begin{cases} 2\hat{\Theta}_A s(z_i, \hat{\beta}_u) & 1 \leq i \leq m \\ -2\hat{\Theta}_B s(z_i, \hat{\beta}_u) & m+1 \leq i \leq n. \end{cases} \quad (4.5)$$

Point of departure from the mentioned work is that in many statistical models (where our general methodology applies), the sequence $\{\hat{R}_1, \ldots, \hat{R}_n\}$ is not independent and identically distributed. Under certain regularity conditions, we shall show that estimates $\widehat{R}_i$ are sufficiently good such that the resampled distribution of $T_n^{BS}$ offers a good approximation of the distribution of $T_n$. Hence, we can take as the critical value for $T_n$ the $(1 - \alpha)$-quantile of $T_n^{BS}$ (while holding the observations fixed), denoted by $\mathcal{Q}(1 - \alpha, T_n^{BS})$, where $\alpha$ is the nominal size of the test.

With these steps in place, we are ready to define our procedure in Algorithm 1.

---

**Algorithm 1** Projection pursuit testing (PPTest)

---

**Require:** Observations $\{z_1, \ldots, z_n\}$ and a test size $\alpha \in (0, 1)$
**Ensure:** For a test with nominal size $\alpha$, implement the following
1: Compute the initial estimator $\hat{\beta}_u$
2: Compute the projection pursuit estimator

$$\hat{\beta}_d = \arg\min_{\beta \in \mathbb{R}^p} \left\{ \|\beta - \hat{\beta}_u\|_1 , \text{ s.t. } \beta \in \mathcal{B}_0 \right\}.$$

3: Compute $\hat{R}_1, \cdots, \hat{R}_n$ as in (4.5)
4: Compute the test statistic $T_n = \sqrt{n} \max_{1 \leq j \leq p} \left| \hat{\beta}_{u,j} - \hat{\beta}_{d,j} - \hat{\delta}_j \right|$ as in (4.1).
5: **for** $b = 1, \cdots, B$ **do**
6:     Generate $\{\xi_i\}_{i=1}^n$ a sequence of i.i.d $N(0, 1)$ that are also independent of $z_1, \ldots, z_n$
7:     Compute bootstrap test statistic $T_{n,b} = T_n^{BS}$ as in (4.4).
8: **end for**
9: Set $\mathcal{Q}(1 - \alpha, T_n^{BS})$ as $(1 - \alpha)$ quantile of the sequence $T_{n,1}, T_{n,2}, \ldots, T_{n,B}$.
    **return** Reject $H_0$ if and only if $T_n > q_{1-\alpha}$.

---

## 4.2 Theoretical properties

In this section we outline theoretical guarantees for the Projection Pursuit Test. In order to encompass various regression models we impose high-level assumptions on the vectors $R_i$ as defined in (4.3). We have seen that these assumptions for the case of linear models lead to extremely weak conditions. Moreover, in Section 5 we show that they also lead to weak conditions for the case of logistic regression.

### 4.2.1 Size control

In this subsection, we study the size properties of the proposed test. We begin by outlining the conditions imposed on the vectors $R_i \in \mathbb{R}^p$, (4.3).

**Assumption 1.** *There exists a $\sigma$-algebra $\mathcal{F}_n$, such that $\hat{\delta}$ and $R_i$ satisfy (i) $\{R_i\}_{i=1}^n$ are mean zero random vectors that, conditional on $\mathcal{F}_n$ are independent for all $i$; (ii) $\max_{1 \leq j \leq p} \sum_{i=1}^n \left| R_{i,j}^2 - \mathbb{E}(R_{i,j}^2 \mid \mathcal{F}_n) \right| = o_P(n)$ and $\max_{1 \leq j \leq p} |\sum_{i=1}^n R_{i,j}| = o_P(n)$; (iii) There exists a constant $b > 0$ such that $\lim_{n \to \infty} \mathbb{P}\left( \min_{1 \leq j \leq p} \sum_{i=1}^n \mathbb{E}(R_{i,j}^2 \mid \mathcal{F}_n) > bn \right) = 1$; (iv) There exists an $\mathcal{F}_n$-measurable positive random variable $B_n$, such that $B_n^2 \log^5(p \vee n)/n = o_P(1)$ and almost surely $\max_{1 \leq j \leq p} \sum_{i=1}^n \mathbb{E}(|R_{i,j}|^3 \mid \mathcal{F}_n) \leq nB_n$, $\max_{1 \leq j \leq p} \sum_{i=1}^n \mathbb{E}(|R_{i,j}|^4 \mid \mathcal{F}_n) \leq nB_n^2$ and $\max_{1 \leq i \leq n, \, 1 \leq j \leq p} \mathbb{E}(\exp(|R_{i,j}|/B_n) \mid \mathcal{F}_n) \leq 2$.*

Assumption 1(i)-(v) guarantee the validity of the bootstrap procedure. Assumptions 1(i)-(ii) state that the vectors $R_i$ are centered and should concentrate well enough around its mean. They are trivially satisfied for sub-exponential random vectors, for example. Assumption 1(iii) rules out asymptotically vanishing variance in components of $R_i$ and Assumption 1(iv) aims to guarantee that $R_i$'s do not have extreme values that can dominate their partial sum.

Next, we present conditions needed to hold for the candidate estimates $\hat{\Theta}_A$, $\hat{\Theta}_B$ of the precision matrix; they specify the quality of the estimators $\hat{\beta}_u$, $\hat{\beta}_d$, $\hat{\Theta}_A$ and $\hat{\Theta}_B$ – a large class of sparsity-encouraging estimators satisfy these conditions and allow PPTest to be a consistent test.

**Assumption 2.** *Suppose that, under the null hypothesis $H_0$, there exist sequences of positive constants $\lambda_{1,n}, \lambda_{2,n} = o(\sqrt{n^{-1} \log p})$ such that (i) $\sup_{t \in [0,1]} \|[\mathbb{I}_p - \hat{\Theta}_A \hat{H}_A(\hat{\beta}_u - t(\hat{\beta}_u - \beta_*))](\hat{\beta}_u - \beta_*)\|_\infty = O_P(\lambda_{1,n})$ and $\sup_{t \in [0,1]} \|[\mathbb{I}_p - \hat{\Theta}_B \hat{H}_B(\hat{\beta}_d - t(\hat{\beta}_d - \beta_*))](\hat{\beta}_d - \beta_*)\|_\infty = O_P(\lambda_{1,n})$; (ii) For the $\sigma$-algebra $\mathcal{F}_n$ defined in Assumption 1, there exist $\mathcal{F}_n$-measurable matrices $\Theta_A$ and $\Theta_B$ such that $\|(\hat{\Theta}_A - \Theta_A) \sum_{i=1}^m s(z_i, \beta_*)\|_\infty / m = O_P(\lambda_{2,n})$ and $\|(\hat{\Theta}_B - \Theta_B) \sum_{i=m+1}^n s(z_i, \beta_*)\|_\infty / m = O_P(\lambda_{2,n})$. Moreover, $\|\hat{\beta}_u - \beta_*\|_1 = o_P(s\sqrt{\log p/n})$.*

Note that the above conditions, although presented at the high-level, are quite mild. Assumption 2 (i) is related to the stability of the Hessian matrix and its estimator. In Section 3 and Section 5, show that these are satisfied in linear and logistic linear models, respectively, with $\lambda_{1,n}$ of the order of $o(\sqrt{n^{-1}\log p})$ without many other model restrictions. Assumption 2 (ii) states certain consistency condition for $\hat{\Theta}_A$ and $\hat{\Theta}_B$ (if suitably chosen) together with the average of the residual of the loss function $L$. For the case of linear models with least squares loss, this condition disappears as we can choose $\Theta_A = \hat{\Theta}_A$ and similarly $\Theta_B = \hat{\Theta}_B$. However, for non-linear models, this condition is satisfied, for $\Theta_A = \Theta_B = [\nabla^2_\beta \mathbb{E} l(z, \beta_*)]^{-1}$ and as long as $\|\Theta_A^{-1}\hat{\Theta}_A - \mathbb{I}_p\|_1 = o_P(1)$ and $s$ has bounded third moment. The later two are satisfied for a large number of models; see Van de Geer et al. (2014) and Javanmard and Montanari (2014). In particular, these assumptions are not stronger than those of Ning and Liu (2014). However, unlike these works, we consider a much broader and more difficult/complex null hypothesis – for example, the null can be in full $p$ dimensional space.

**Theorem 3.** *Consider Algorithm 1. Let $\log p = o(\sqrt{n})$ and $s = o(n^{1/4}/\sqrt{\log p})$. Assume that the estimating sequence $\hat{R}_i$ is such that $\max_{1 \leq j \leq p} \sum_{i=1}^n (\hat{R}_{i,j} - R_{i,j})^2 = o_P(n/\log^2(p \vee n))$. Moreover, let Assumptions 1 and 2 hold. Then, under the null hypothesis (1.2) $H_0 : \beta_* \in \mathcal{B}_0$, the test statistic $T_n$ is asymptotically valid in that*

$$\limsup_{n \to \infty} \sup_{\alpha \in (0,1)} \left| \mathbb{P}\left(T_n > \mathcal{Q}(1 - \alpha, T_n^{BS})\right) - \alpha \right| = 0.$$

Theorem 3 establishes the consistency of the proposed method. It says that Algorithm 1 provides a testing procedure with asymptotically exact control of the size, independent of the structure of the set $\mathcal{B}_0$. The approximation result above is obtained by developing new high-dimensional bootstrap results; the bootstrap is based on dependent sequences whose correlations depends on the regression model and needs to hold irrespective of the null set. As a result the proof is quite involved.

*Remark* 1. Given the generality of the structure of the null hypothesis, this result is quite remarkable. Typically, proving asymptotically exact size control involves deriving the asymptotic distribution of the test statistic. Even for tests of constraints for low-dimensional $\beta_*$, such as the classical Wald tests, likelihood ratio tests and score tests, formulating the asymptotic distribution of the test statistics requires regularity conditions on the constraint, e.g. differentiability and non-singular Jacobian matrix of the function representing the constraint; see Chapter 12 in Lehmann and Romano (2006) or Chapter 6 in Shao (2003). In contrast, the size of our test is asymptotically exact for *any* hypothesis on $\beta_*$, which can have a dimension much larger than the sample size.

### 4.2.2 Power control

In this subsection, we discuss the power of the test. For a given sequence of positive numbers $c_n$ we define a class of parameter spaces

$$\mathcal{B}_1(c_n) = \left\{ \beta \in \mathbb{R}^p : \min_{\delta \in \mathcal{B}_0} \|\delta - \beta\|_\infty \geq c_n \right\}.$$

If $\beta_* \in \mathcal{B}_1(c_n)$ then at least one element of $\beta_*$ lies outside the null set $\mathcal{B}_0$.

Note that deriving the power properties for a general set $\mathcal{B}_0$ is extremely challenging. Even for well-studied examples of $\mathcal{B}_0$, theoretical results regarding power properties can be extremely difficult to establish. Consider, for example, the multiple testing case, where the test is required to control the family-wise error rate (FWER) or false discovery rate (FDR). Many of the multiple testing problems can be cast as testing $\beta_* \in \mathcal{B}_0$, where $\mathcal{B}_0 = \{\beta \mid \beta_J = 0\}$ and $J$ is a known subset of $\{1, \cdots, p\}$; see Fan et al. (2012); Bühlmann (2013); Barber and Candès (2015) for example. The vast majority of the literature in this area is concerned with the size properties in terms of FWER or FDR, while very little work, except on the consistency of the tests, has been done regarding the power properties, such as optimality and local power analysis. A careful characterization of the structure of the test statistic under the alternative hypothesis is needed for such analysis and is typically only possible for a small class of alternatives.

Despite these difficulties, we can establish some theoretical results on the power of the proposed projection pursuit test without considering the structure of the set $\mathcal{B}_0$. Our result also has the flavor of local power analysis because we allow $c_n$ to decay to zero.

**Theorem 4.** *Consider Algorithm 1 with a nominal size $\alpha \in (0,1)$. Let the assumptions in Theorem 3 hold. Suppose that $c_n \gg n^{-1/4}$. Then for any fixed $a > 0$, we have that under $H_{1,n}$,*

$$\lim_{n \to \infty} \inf_{\beta_* \in \mathcal{B}_1(c_n)} \mathbb{P}(T_n > \mathcal{Q}(1 - \alpha, T_n^{BS})) = 1.$$

A few comments are in order. It can be seen from Theorem 4 that it only requires one of the entries $\beta_*$ to lie outside of the null set $\mathcal{B}_0$ with distance bigger than $O(n^{-1/4})$ for the test to correctly reject $H_0 : \beta_* \in \mathcal{B}_0$. Also the result holds true uniformly over a large class of sets $\mathcal{B}_0$ – we only require it to be a closed set in $\mathbb{R}^p$. Observe that the test depends on the null set $\mathcal{B}_0$ directly. However, power is independent of the "difficulty" of the null set $\mathcal{B}_0$. In that sense PPTest is robust in the sense that its properties do not change with the changes in $\mathcal{B}_0$.

Rate optimality of PPTest is very difficult to verify. Although minimax theory has had some success in high-dimensional models, the development of a satisfactory theory for general models or complex hypothesis has proved to be extremely difficult. To the best of our knowledge there exists no results that cover estimation of a general functional $\psi(\beta_*)$ – existing results for $\beta_*$ extend to $\psi(\beta_*)$ only in a few special cases and are model specific. Naively, we can see that under a class of alternatives $\mathcal{B}_1(c_n)$, the tests based on de-biased estimators would suffer suboptimal power – see Cai and Guo (2015) where authors document that $a^\top \beta_*$ with dense vectors $a$ cannot be tested in linear models without explicit knowledge of $\|\beta_*\|_0$. Hence, in this way, our test and the rate of Theorem 4 serve as unique benchmarks. For this reason, our test can be used in an extremely wide range of setups.

## 5 Applications to logistic regression

In Sections 4 we have introduced a general framework for testing $H_0$ (1.1) and simple constructions that can be applied to a wide range of models. In this section, we illustrate that the high-level conditions of Section 4 are very mild and in fact match those of testing single entries of $\beta_*$ (a much simpler problem than ours) – and are hence in this sense perhaps optimal.

Consider a logistic regression model, where the i.i.d. observations $\{(y_i, x_i)\}_{i=1}^n$, with $y_i \in \{0, 1\}$, follow

$$P(y_i = 1 \mid x_i) = \exp(x_i^\top \beta_*)/[1 + \exp(x_i^\top \beta_*)]. \tag{5.1}$$

Under the framework (1.1), we have $z_i = (y_i, x_i)$ and $l(z_i, \beta) = -y_i x_i^\top \beta + b(x_i^\top \beta)$ with $b(u) = \log(1 + \exp(u))$. The true parameter value is defined by

$$\beta_* = \arg\min_{\beta \in \mathbb{R}^p} L(\beta) \qquad \text{with } L(\beta) = \mathbb{E} l(z_1, \beta). \tag{5.2}$$

The construction in Section 4 demands a construction of two estimates. The initial one, can be taken to be a $\ell_1$-penalized logistic estimator

$$\hat{\beta}_u = \arg\min_{\beta \in \mathbb{R}^p} \left\{ n^{-1} \sum_{i=1}^n [-y_i x_i^\top \beta + \log(1 + \exp(x_i^\top \beta))] + \lambda \|\beta\|_1 \right\}, \tag{5.3}$$

for $\lambda > 0$ chosen as $\lambda \asymp \sqrt{n^{-1} \log p}$. Then, we construct a $l_1$ projection pursuit estimate

$$\hat{\beta}_d = \arg\min_{\beta \in \mathbb{R}^p} \left\{ \|\beta - \hat{\beta}_u\|_1, \text{ s.t. } \beta \in \mathcal{B}_0 \right\}. \tag{5.4}$$

We then move on to construct an estimate of the bias of the two regularized estimators. Following the proposal presented in Section 4 we define

$$\hat{\delta}_{SP} = \hat{\Theta}_A m^{-1} \sum_{i=1}^{m} x_i[-y_i + b'(x_i^\top \hat{\beta}_u)] - \hat{\Theta}_B m^{-1} \sum_{i=m+1}^{n} x_i[-y_i + b'(x_i^\top \hat{\beta}_d)] \quad (5.5)$$

where $\hat{\Theta}_A$ and $\hat{\Theta}_B$ are candidate estimates of the precision matrix $\Theta_* := [\mathbb{E}\nabla_\beta^2 l(z_1, \beta_*)]^{-1} = [\mathbb{E}x_1 x_1^\top b''(y_1, x_1^\top \beta_*)]^{-1} \in \mathbb{R}^{p \times p}$. For estimating $\Theta_*$, we use the node-wise Lasso estimator proposed by Van de Geer et al. (2014). For that end, it is important to notice that the two contrasting estimators are de-biased on two subsamples independently. Recall that $m = n/2$. Let $\hat{U}_A = (\hat{U}_{A,1}, \cdots, \hat{U}_{A,p}) \in \mathbb{R}^{m \times p}$ with

$$\hat{U}_{A,j} = \left(x_{1,j}\sqrt{b''(x_1^\top \hat{\beta}_u)}, \cdots, x_{m,j}\sqrt{b''(x_m^\top \hat{\beta}_u)}\right)^\top \in \mathbb{R}^m.$$

and $\hat{U}_B = (\hat{U}_{B,1}, \cdots, \hat{U}_{B,p}) \in \mathbb{R}^{m \times p}$ with

$$\hat{U}_{B,j} = \left(x_{m+1,j}\sqrt{b''(x_{m+1}^\top \hat{\beta}_u)}, \cdots, x_{n,j}\sqrt{b''(x_n^\top \hat{\beta}_u)}\right)^\top \in \mathbb{R}^m.$$

Then, for $1 \leq j \leq p$ and $\eta \asymp n^{-1/2} \log p$, compute

$$\begin{cases} \hat{\gamma}_{A,j} = \underset{\gamma \in \mathbb{R}^{p-1}}{\arg\min} \left\{ (2m)^{-1} \|\hat{U}_{A,j} - \hat{U}_{-A,j}\gamma\|_2^2 + \eta\|\gamma\|_1 \right\} \\ \hat{\gamma}_{B,j} = \underset{\gamma \in \mathbb{R}^{p-1}}{\arg\min} \left\{ (2m)^{-1} \|\hat{U}_{B,j} - \hat{U}_{B,-j}\gamma\|_2^2 + \eta\|\gamma\|_1 \right\}, \end{cases} \quad (5.6)$$

and set $\hat{\Theta}_A = (\hat{\Theta}_{A,1}, \cdots, \hat{\Theta}_{A,p})^\top$ and $\hat{\Theta}_B = (\hat{\Theta}_{B,1}, \cdots, \hat{\Theta}_{B,p})^\top$ as follows

$$\begin{cases} \hat{\Theta}_{A,j,j} = [m^{-1}\hat{U}_{A,j}^\top(\hat{U}_{A,j} - \hat{U}_{A,-j}\hat{\gamma}_{A,j})]^{-1} \text{ and } \hat{\Theta}_{A,j,-j} = -\hat{\Theta}_{A,j,j}\hat{\gamma}_{A,j} & \text{for } 1 \leq j \leq p \\ \hat{\Theta}_{B,j,j} = [m^{-1}\hat{U}_{B,j}^\top(\hat{U}_{B,j} - \hat{U}_{B,-j}\hat{\gamma}_{B,j})]^{-1} \text{ and } \hat{\Theta}_{B,j,-j} = -\hat{\Theta}_{B,j,j}\hat{\gamma}_{B,j} & \text{for } 1 \leq j \leq p. \end{cases} \quad (5.7)$$

The tuning parameter can be chosen as $\eta \asymp n^{-1/2} \log p$. As an alternative to the Lasso-type of estimators for $\hat{\Theta}_A$ and $\hat{\Theta}_B$, one may use a CLIME-type estimator by imposing the constraint $\|\hat{U}_{A,j}^\top \hat{U}_A \Theta_j/m - e_j\|_\infty \leq \bar{\eta}$, while minimizing $\|\Theta_j\|_1$, where the minimizer $\Theta_j$ is the estimate for the $j$th column of $\Theta_*$. This estimator would require a tuning parameter $\bar{\eta}$ to depend on the rate of convergence of the initial estimator $\hat{\beta}_u$. Since the rate of convergence typically depends on the sparsity level $\|\beta_*\|_0$, which is rarely known, CLIME-type estimators are less desirable in our setup. Here, we differ from the case of linear models, in that $\hat{\Theta}_A \neq \hat{\Theta}_B$. This construction offers convenience in technicalities of the proof and delivers weaker regularity conditions.

Then, we have all the elements to define the test statistic

$$T_n = \sqrt{n}\|\hat{\beta}_u - \hat{\beta}_d - \hat{\delta}\|_\infty.$$

For computing its critical value, we compute the bootstrap test statistic

$$T_n^{BS} = \max_{1 \leq j \leq p} n^{-1/2} \left| \sum_{i=1}^n (\widehat{R}_{i,j} - R_j^*)\xi_i \right|,$$

where $\{\xi_i\}_{i=1}^n$ are i.i.d standard normal random variables and following Section 4 we define

$$\widehat{R}_i = \begin{cases} -2\hat{\Theta}_A x_i[-y_i + b'(x_i^\top \hat{\beta}_u)] & \text{for } 1 \leq i \leq m \\ 2\hat{\Theta}_B x_i[-y_i + b'(x_i^\top \hat{\beta}_u)] & \text{for } m+1 \leq i \leq n \end{cases} \quad \text{and} \quad R^* = n^{-1} \sum_{i=1}^n \widehat{R}_i. \quad (5.8)$$

Observe that the non-linearity of the model presents theoretical challenges, but not methodological ones. We then reject the null $H_0$ (1.2) if $T_n > q_n(1-\alpha)$. Despite high-dimensionality, non-linearity and various dependence structure, we are able to obtain the next result confirming asymptotic Type I error control of the above test procedure.

**Theorem 5.** *Consider the model (5.1). Let there exist constants $\kappa_1, \kappa_2 \in (0, \infty)$, such that the eigenvalues of $\Sigma_X = \mathbb{E}x_1 x_1^\top$ and $\Theta_* = [\mathbb{E}x_1 x_1^\top b''(y_1, x_1^\top \beta_*)]^{-1}$ lie in $[\kappa_1, \kappa_2]$, where $b''(u) = \exp(u)/[1+\exp(u)]^2$ Additionally, let the design $x_i = (1, z_i^\top)^\top \in \mathbb{R}^p$ is such that $\mathbb{E}z_i = 0$ and there exists a constant $\kappa_3 \in (0, \infty)$ such that $\|\Sigma_X^{-1/2} x_1\|_{\psi_2} \leq \kappa_3$. Suppose that $\log p = o(n^{1/5})$,*

$$\|\beta_*\|_0 = o\left(n^{1/4}/\log^{5/4}(p \vee n)\right) \quad \text{and} \quad \max_{1 \leq j \leq p} \|\Theta_{*,j}\|_0 = o\left(\sqrt{n}/\log(p \vee n)\right).$$

*Then*

$$\limsup_{n \to \infty} \sup_{\alpha \in (0,1)} \left| \mathbb{P}\left(T_n > \mathcal{Q}(1-\alpha, T_n^{BS})\right) - \alpha \right| = 0.$$

A few comments are needed. Assumptions utilized in Theorem 5 are very similar to the regularity conditions commonly imposed for generalized linear models; see Theorem 3.3 in Van de Geer et al. (2014) although we attack a more difficult problem than that of univariate testing; one special case is simultaneous testing with growing number of tests or testing for signal to noise ratio in logistic regression.

*Remark* 2. Notice that the sparsity requirement for $\beta_*$ is $\|\beta_*\|_0 = o\left(n^{1/4}/\log^{5/4}(p \vee n)\right)$ appears stronger than the usual requirement of $\|\beta_*\|_0 = o\left(n^{1/2}/\log^c(p \vee n)\right)$ for some constant $c > 0$. This stronger condition on the sparsity can be viewed as the price we pay for allowing for general $\mathcal{B}_0$. However, condition on the sparsity of the precision matrix is the same as those of linear models.

Moreover, it is worth pointing that with minor changes in the proof, we can allow for general convex functions $b(\cdot)$ and establish the same results as in Theorem 5 for generalized linear models. However, this typically requires additional restrictive assumptions on the design and link functions and estimations, when the support of $y_i$ is unbounded, hence we do not provide further details.

**Theorem 6.** *Consider the model (5.1). Let the assumptions in Theorem 5 hold. Suppose that* $\min_{\beta \in \mathcal{B}_0} \|\beta - \beta_*\|_\infty \geq c_n$ *with* $c_n \gg n^{-1/4}$. *Then we have that*

$$\lim_{n \to \infty} \mathbb{P}(T_n > \mathcal{Q}(1 - \alpha, T_n^{BS})) = 1.$$

The result above, although a corollary of Theorem 4 presents a unique power guarantees for generalized linear models in high-dimensions. Existing work only covers testing single elements of $\beta_*$. Some attempts at simultaneous testing exists (e.g. Ning and Liu (2014)) but no power guarantees are provided. Hence, even for simultaneous testing, Theorem 6 implies that our test provides novel guarantees.

## 6 Numerical Examples

This section examines the finite-sample performance of the proposed method. We consider both the linear regression model (3.1) and the logistic regression model (5.1). For both models, the design matrix $X$ has i.i.d rows generated from $x_i \sim N(0, \Sigma_X)$ with Toeplitz matrix $\Sigma_{X,i,j} = \rho^{|i-j|}$ and the model parameter $\beta_* = (1, \cdots, 1, 0, \cdots, 0)^\top \in \mathbb{R}^p$ with $\|\beta_*\|_0 = 4$. In the linear model, we generate $\varepsilon \sim N(0, \mathbb{I}_n)$ and $Y = X\beta + \varepsilon$; for the logistic regression model, $y = (y_1, \cdots, y_n)^\top$ and $y_i = \mathbf{1}\{u_i \leq \exp(x_i^\top \beta_*)/[1 + \exp(x_i^\top \beta_*)]\}$, where $u_i$ is independent of $X$ and is drawn from the uniform distribution on the interval $(0, 1)$. We set $n = 200$ and compute the rejection probabilities based on 100 random samples. We consider three hypotheses:

(a) $H_0^{(A)} : \|\beta_*\|_0 \leq s_0$

(b) $H_0^{(B)} : \min_{j \in \text{supp}(\beta_*)} |\beta_{*,j}| \geq r_0$

(c) $H_0^{(C)} : \|\beta_*\|_2 \leq c_0$.

The initial estimators $\hat{\beta}_u$ are computed as follows. For the linear model, we use the scaled-Lasso (Sun and Zhang, 2012) with universal tuning parameter; for the logistic model, we use the Lasso estimator with tuning parameter chosen by 10-fold cross validation.

The project pursuit estimators $\hat{\beta}_d$ are defined as follows. For hypothesis $H_0^{(A)}$, let $\hat{J} \subset \{1,...,p\}$ be the indexes of the $s_0$ largest entries in magnitude of $\hat{\beta}_d$, i.e., $|\hat{J}| = s_0$ and $\hat{\beta}_{u,j_1} \geq \hat{\beta}_{u,j_2}$ for any $j_1 \in \hat{J}$ and $j_2 \notin \hat{J}$. Then we define $\hat{\beta}_d$ as $\hat{\beta}_{d,j} = \hat{\beta}_{d,j} \mathbf{1}\{j \in \hat{J}\}$. For hypothesis $H_0^{(B)}$, $\hat{\beta}_{d,j} = \rho(\hat{\beta}_{u,j}, r_0)$ for all $1 \leq j \leq p$, where $\rho(\cdot,\cdot)$ is defined in Lemma 3. For $H_0^{(C)}$, define $t \mapsto a(t)$ by $a(t) = \arg\min_{v \in \mathbb{R}^p} \|\hat{\beta}_u - v\|_2^2 + t\|v\|_1$ and $t^* = \arg\min_{t \geq 0} \|a(t)\|_1$ s.t. $\|\hat{\beta}_u + a(t)\|_2 \leq c_0$. Then $\hat{\beta}_d = a(t^*) + \hat{\beta}_u$.

Table 1 summarizes the numerical performance of the projection pursuit test for three hypothesis above. We set $s_0 = 4$, $r_0 = 1$ and $c_0 = 2$ and report the average Type I error rates. We can see that the projection pursuit test effectively deals with both the convex and non-convex null sets $\mathcal{B}_0$, and performs very closely to the nominal size $\alpha$. Moreover, the tests show stability across the dimension $p$ and the correlation of the design matrix. This numeric evidence is consistent with theoretical results presented in Sections 3-5.

Table 1: Size properties

| | Linear regression model | | | | | | | | | | | |
|---|---|---|---|---|---|---|---|---|---|---|---|---|
| | $H_0 : \|\beta_*\|_0 \leq s_0$ | | | | $H_0 : \min_{j \in \text{supp}(\beta_*)} |\beta_{*,j}| \geq r_0$ | | | | $H_0 : \|\beta_*\|_2 \leq c_0$ | | | |
| $p \backslash \rho$ | 0 | 0.25 | 0.50 | 0.75 | 0 | 0.25 | 0.50 | 0.75 | 0 | 0.25 | 0.50 | 0.75 |
| 200 | 0.04 | 0.02 | 0.03 | 0.05 | 0.03 | 0.02 | 0.03 | 0.01 | 0.06 | 0.05 | 0.06 | 0.05 |
| 350 | 0.04 | 0.03 | 0.02 | 0.04 | 0.01 | 0.03 | 0.04 | 0.02 | 0.07 | 0.05 | 0.06 | 0.03 |
| 500 | 0.03 | 0.05 | 0.03 | 0.04 | 0.01 | 0.03 | 0.04 | 0.02 | 0.05 | 0.05 | 0.06 | 0.03 |
| | Logistic regression model | | | | | | | | | | | |
| | $H_0 : \|\beta_*\|_0 \leq s_0$ | | | | $H_0 : \min_{j \in \text{supp}(\beta_*)} |\beta_{*,j}| \geq r_0$ | | | | $H_0 : \|\beta_*\|_2 \leq c_0$ | | | |
| $p \backslash \rho$ | 0 | 0.25 | 0.50 | 0.75 | 0 | 0.25 | 0.50 | 0.75 | 0 | 0.25 | 0.50 | 0.75 |
| 200 | 0.04 | 0.02 | 0.03 | 0.05 | 0.03 | 0.02 | 0.03 | 0.01 | 0.06 | 0.05 | 0.06 | 0.05 |
| 350 | 0.04 | 0.03 | 0.02 | 0.04 | 0.01 | 0.03 | 0.04 | 0.02 | 0.07 | 0.05 | 0.06 | 0.03 |
| 500 | 0.03 | 0.05 | 0.03 | 0.04 | 0.01 | 0.03 | 0.04 | 0.02 | 0.05 | 0.05 | 0.06 | 0.03 |

Next, we compare the performance of the projection pursuit test in detecting alternative hypothesis corresponding to the nulls $H_0^{(A)}$, $H_0^{(B)}$ and $H_0^{(C)}$. In each of the 100 replications we simulate the Toeplitz design setting with correlation $\rho = 0.5$ and sample size $p = 500$. The averages of the rejection probabilities are collected in in Table 2. This table shows that the proposed PPTests reach power

Table 2: Power properties

| | $H_0: \|\beta_*\|_0 \leq s_0$ | | | $H_0: \min_{j \in \text{supp}(\beta_*)} |\beta_{*,j}| \geq r_0$ | | | $H_0: \|\beta_*\|_2 \leq c_0$ | |
|---|---|---|---|---|---|---|---|---|
| $s_0$ | Linear | Logit | $r_0$ | Linear | Logit | $c_0$ | Linear | Logit |
| 4 | 0.05 | 0.09 | 1.0 | 0.06 | 0.06 | 2.0 | 0.04 | 0.08 |
| 3 | 0.99 | 0.19 | 1.2 | 0.23 | 0.18 | 1.2 | 0.37 | 0.40 |
| 2 | 0.99 | 0.58 | 1.4 | 0.54 | 0.48 | 1.0 | 0.76 | 1.00 |
| 1 | 1.00 | 1.00 | 1.6 | 0.95 | 0.78 | 0.9 | 0.91 | 1.00 |

against all the alternatives in both linear and logistic model.

# 7 Conclusion and discussion

In this article, we propose a unified method for testing a widespread collection of hypotheses of regression parameters in sparse high-dimensional models. The methodology is centered around a newly introduced high-dimensional projection pursuit estimator. As a hypothesis-driven projection of the initial estimator, projection pursuit estimator is the estimator closest to the initial one that is simultaneously forced to satisfy the null hypothesis. In this way the geometry of the null set is directly embedded in the estimator and further in the test statistic, making it highly robust to the specifications of the null.

It is worth emphasizing that this estimator and the projection pursuit test are valid even in low-dimensional problems and provide new inference principles. Without imposing any constraint on the structure of the null hypothesis, the proposed projection pursuit inference method is with asymptotically exact size control for a broad spectrum of high-dimensional models. Therefore, this work widens the scope of the state-of-the-art methodology. For the first time, we were able to test the sparsity of the model and the beta-min condition of the model parameters for example. Moreover, the proposed methodology opens doors to many unexplored areas in high-dimensional inferential statistics. Examples include tests of monotonicity, unimodality or concavity of high-dimensional regression functions.

Additionally, the proposed methodology leaves open numerous avenues for further study. Although in our work we focused on a specific sparsity structure (defined through the $\ell_0$-norm of the model parameter), our methodology extends far beyond this kind of sparsity considerations. Namely, group or hierarchical sparsity structures are particularly important in applications and our test opens the door to designing tests suitable for these models. Moreover, although in this work we consider a statistic that is linear in the $R_i$, the framework and ideas of this work are not intimately tied to this formulation.

Finally, our test provides power guarantees for an extremely large set of problems in high-dimensions. For many testing procedures it is extremely difficult, if not impossible, to establish theoretical results regarding optimality/efficiency. However, due to its direct utilization of the set $\mathcal{B}_0$, we believe that it opens the door to many open problems in the minimax theory of estimation in high-dimensions and serves as a benchmark when no other tests are available. We shall leave these unexplored issues to future research.

# Supplementary Materials

Here, we provide details of all of the theoretical statements made in the main document. We first show the results in Section 2. Then the proof of master theorems in Section 4 will be provided. We continue by proving the results in Sections 3 and 5. Lastly, we provide details of the proofs of a number of lemmas and technical tools.

## A  Proof of results in Section 2

**Proof of Lemma 1.** By the triangular inequality, $\|\hat{\beta}_d - \beta_*\|_1 \leq \|\hat{\beta}_u - \beta_*\|_1 + \|\hat{\beta}_u - \hat{\beta}_d\|_1$. Under $H_0$, $\beta_* \in \mathcal{B}_0$ and thus $\|\hat{\beta}_d - \hat{\beta}_u\|_1 \leq \|\beta_* - \hat{\beta}_u\|_1$. The desired result follows. $\square$

**Proof of Lemma 2.** Fix an arbitrary $u \in \mathbb{R}^p$ with $\|u\|_0 \leq s_0$. We have that

$$\|u - v\|_1 = \sum_{j=1}^{p} |u_{\pi(j)} - v_{\pi(j)}|$$

$$= \sum_{j \in \text{supp}(u)} |u_{\pi(j)} - v_{\pi(j)}| + \sum_{j \notin \text{supp}(u)} |v_{\pi(j)}|$$

$$\geq \sum_{j \notin \text{supp}(u)} |v_{\pi(j)}| \stackrel{(i)}{\geq} \min_{J \subseteq \{1,\cdots,p\}\ |J| \geq p-s_0} \sum_{j \in J} |v_{\pi(j)}| \stackrel{(ii)}{=} \sum_{j=s_0+1}^{p} |v_{\pi(j)}|,$$

where $(i)$ follows by the fact that $|\{1, \cdots, p\} \setminus \text{supp}(u)| \geq p - s_0$ and $(ii)$ follows by the ranking property of $\pi$ (i.e., $|v_{\pi(1)}| \geq \cdots \geq |v_{\pi(p)}|$). Since $u$ is arbitrary, we have

$$\min_{\|\beta\|_0 \leq s_0} \|\beta - v\|_1 \geq \sum_{j=s_0+1}^{p} |v_{\pi(j)}| = \|\tilde{v} - v\|_1.$$

Since $\|\tilde{v}\|_0 \leq s_0$, the desired result follows. $\square$

**Proof of Lemma 3.** Let $S = \arg\min_{\beta \in \mathbb{R}^p} \|\beta - v\|_1$ s.t. $\min_{j \in \text{supp}(\beta)} |\beta_j| \geq c$. Fix any $j_0 \in \{1, \cdots, p\}$ and $\hat{v} \in S$. Define $\dot{v} \in \mathbb{R}^p$ with $\dot{v}_k = \hat{v}_k$ for $k \neq j_0$ and $\dot{v}_{j_0} = \rho(v_{j_0}, c)$. It is not hard to see that

$$|\dot{v}_{j_0} - v_{j_0}| = \inf_{|x| \in \{0\} \bigcup [c, \infty)} |x - v_{j_0}|.$$

Thus, $\|\hat{v} - v\|_1 - \|\dot{v} - v\|_1 = |\hat{v}_{j_0} - v_{j_0}| - |\dot{v}_{j_0} - v_{j_0}| \geq 0$. Since $\min_{j \in \text{supp}(\dot{v})} |\dot{v}_j| \geq c$, we have $\dot{v} \in S$.

Hence, $\forall \hat{v} \in S$ and $\forall j_0 \in \{1, \cdots, p\}$, we have $t_{j_0}(\hat{v}) \in S$, where $\forall u \in \mathbb{R}^p$, $t_{j_0}(u)$ is the vector $u$ with its $j_0$th coordinate set to $\rho(v_{j_0}, c)$. Thus, by induction, $(t_1 \circ t_2 \circ \cdots \circ t_p)(\hat{v}) \in S$. The proof is complete since $\tilde{v} = (t_1 \circ t_2 \circ \cdots \circ t_p)(\hat{v}) \ \forall \hat{v} \in \mathbb{R}^p$. □

**Proof of Lemma 4.** By introducing the Lagrangian multiplier, we have that, for some constant $\lambda > 0$,

$$\underset{\|Q\beta\|_2^2 \leq c^2}{\arg\min} \|\beta - v\|_1 = \underset{\beta \in \mathbb{R}^p}{\arg\min} \left[\|\beta - v\|_1 + \lambda(\|Q\beta\|_2^2 - c)\right] = \underset{\beta \in \mathbb{R}^p}{\arg\min} \left[\frac{1}{\lambda}\|\beta - v\|_1 + \|Q\beta\|_2^2\right].$$

By a change of variables, we obtain

$$\underset{\beta \in \mathbb{R}^p}{\arg\min} \left[\frac{1}{\lambda}\|\beta - v\|_1 + \|Q\beta\|_2^2\right] = v + a(1/\lambda).$$

Hence, $\min_{\|Q\beta\|_2^2 \leq c^2} \|\beta - v\|_1 = \|a(t_0)\|_1$ for $t_0 = 1/\lambda$. Notice that $\|Q(v + a(t_0))\|_2 \leq c$. Therefore,

$$\min_{t>0,\ \|Q(v+a(t))\|_2 \leq c} \|a(t)\|_1 \leq \|a(t_0)\|_1 = \min_{\|Q\beta\|_2^2 \leq c^2} \|\beta - v\|_1. \tag{A.1}$$

On the other hand, for $\dot{\beta} = v + a(t_*)$, we have $\|Q\dot{\beta}\|_2 \leq c$ and thus

$$\min_{\|Q\beta\|_2^2 \leq c^2} \|\beta - v\|_1 \leq \|\dot{\beta} - v\|_1 = \|a(t_*)\|_1 = \min_{t>0,\ \|Q(v+a(t))\|_2 \leq c} \|a(t)\|_1. \tag{A.2}$$

It follows, by (A.1) and (A.2), that

$$\min_{t>0,\ \|Q(v+a(t))\|_2 \leq c} \|a(t)\|_1 = \min_{\|Q\beta\|_2^2 \leq c^2} \|\beta - v\|_1.$$

The desired result follows. □

# B  Proof of results in Section 4

## B.1  Proof of Theorem 3

**Lemma 5.** *Suppose that Assumption 2 hold. Let $\hat{\delta}_{SP}$ and $R_i$ be defined in (4.2) and (4.3). Then under $H_0$, we have $\left\|\hat{\beta}_u - \hat{\beta}_d - \hat{\delta}_{SP} - n^{-1}\sum_{i=1}^n R_i\right\|_\infty = O_P(\lambda_{1,n} \vee \lambda_{2,n})$ and $P(\hat{\delta}_{SP} = \hat{\delta}) \to 1$.*

**Proof of Lemma 5.** By the triangular inequality and the definition of $\hat{\delta}_{SP}$ (4.2), we have

$$\left\|\hat{\beta}_u - \hat{\beta}_d - \hat{\delta}_{SP} - n^{-1}\sum_{i=1}^n R_i\right\|_\infty \leq \underbrace{\left\|\hat{\beta}_u - \beta_* - \hat{\Theta}_A m^{-1}\sum_{i=1}^m s(z_i, \hat{\beta}_u) - n^{-1}\sum_{i=1}^m R_i\right\|_\infty}_{J_1}$$

$$+ \left\| \hat{\beta}_d - \beta_* - \hat{\Theta}_B m^{-1} \sum_{i=m+1}^{n} s(z_i, \hat{\beta}_d) + n^{-1} \sum_{i=m+1}^{n} R_i \right\|_\infty. \quad \text{(B.1)}$$

$$\underbrace{\phantom{\left\| \hat{\beta}_d - \beta_* - \hat{\Theta}_B m^{-1} \sum_{i=m+1}^{n} s(z_i, \hat{\beta}_d) + n^{-1} \sum_{i=m+1}^{n} R_i \right\|_\infty}}_{J_2}$$

Notice that

$$
\begin{aligned}
J_1 &= \left\| \hat{\beta}_u - \beta_* - \hat{\Theta}_A m^{-1} \sum_{i=1}^{m} s(z_i, \hat{\beta}_u) + \Theta_A m^{-1} \sum_{i=1}^{m} s(z_i, \beta_*) \right\|_\infty \\
&\leq \left\| \hat{\beta}_u - \beta_* - \hat{\Theta}_A m^{-1} \sum_{i=1}^{m} \left[ s(z_i, \hat{\beta}_u) - s(z_i, \beta_*) \right] \right\|_\infty + \left\| \left( \hat{\Theta}_A - \Theta_A \right) m^{-1} \sum_{i=1}^{m} s(z_i, \beta_*) \right\|_\infty \\
&\stackrel{(i)}{=} \left\| \hat{\beta}_u - \beta_* + \hat{\Theta}_A \int_0^1 \hat{H}_A \left( t(\beta_* - \hat{\beta}_u) + \hat{\beta}_u \right) (\beta_* - \hat{\beta}_u) dt \right\|_\infty + O_P(\lambda_{2,n}) \\
&= \left\| \int_0^1 \left[ \mathbb{I}_p - \hat{\Theta}_A \hat{H}_A \left( t(\beta_* - \hat{\beta}_u) + \hat{\beta}_u \right) \right] (\hat{\beta}_u - \beta_*) dt \right\|_\infty + O_P(\lambda_{2,n}) \\
&\leq \sup_{t \in [0,1]} \left\| \left[ \mathbb{I}_p - \hat{\Theta}_A \hat{H}_A \left( t(\beta_* - \hat{\beta}_u) + \hat{\beta}_u \right) \right] (\hat{\beta}_u - \beta_*) \right\|_\infty + O_P(\lambda_{2,n}) \\
&\stackrel{(ii)}{=} O_P(\lambda_{1,n}) + O_P(\lambda_{2,n}),
\end{aligned}
$$

where $(i)$ holds by Taylor's Theorem (Theorem C.15 of Lee (2012)) at $\beta = \hat{\beta}_u$ and Assumption 2(ii) and $(ii)$ holds by Assumption 2(i). Similarly, we can show that $J_2 = O_P(\lambda_{1,n} + \lambda_{2,n})$. The first claim result follows by (B.1).

We now show the second claim. By the first claim and the assumption that $\lambda_{1,n} \vee \lambda_{2,n} = o(\sqrt{n^{-1} \log p})$, we have that under $H_0$, $\|\hat{\beta}_u - \hat{\beta}_d - \hat{\delta}_{SP} - n^{-1} \sum_{i=1}^{n} R_i\|_\infty = o_P(\sqrt{n^{-1} \log p})$. Hence, under $H_0$, we have

$$
\begin{aligned}
\|\hat{\delta}_{SP}\|_\infty &\leq \|\hat{\beta}_u - \hat{\beta}_d\|_\infty + \left\| n^{-1} \sum_{i=1}^{n} R_i \right\|_\infty + o_P(\sqrt{n^{-1} \log p}) \\
&\stackrel{(i)}{\leq} 3\|\hat{\beta}_u - \beta_*\|_1 + \left\| n^{-1} \sum_{i=1}^{n} R_i \right\|_\infty + o_P(\sqrt{n^{-1} \log p}) = o(n^{-1/4}) + O_P(\sqrt{n^{-1} \log p}) = o_P(n^{-1/4}),
\end{aligned}
$$

where $(i)$ follows by $\|\hat{\beta}_u - \hat{\beta}_d\|_\infty \leq \|\hat{\beta}_u - \hat{\beta}_d\|_1 \leq \|\hat{\beta}_d - \beta_*\|_1 + \|\hat{\beta}_u - \beta_*\|_1$ and Lemma 1. The second claim follows. □

Recall the setup in Theorem 3 as well as in Assumption 1. Let $\mathcal{G}_n$ be the $\sigma$-algebra generated by

$\mathcal{F}_n$ and the data. Define

$$S_R = n^{-1/2} \sum_{i=1}^n R_i, \qquad \tilde{S}_R = n^{-1/2} \sum_{i=1}^n (R_i - \bar{R})\xi_i, \qquad \tilde{S}_{\hat{R}} = n^{-1/2} \sum_{i=1}^n (\hat{R}_i - R^*)\xi_i,$$

where $\bar{R} = n^{-1} \sum_{i=1}^n R_i$ and $R^* = n^{-1} \sum_{i=1}^n \hat{R}_i$. For notational simplicity, we denote $\mathbb{P}(\cdot \mid \mathcal{F}_n)$ and $\mathbb{P}(\cdot \mid \mathcal{G}_n)$ by $\mathbb{P}_{\mathcal{F}_n}(\cdot)$ and $\mathbb{P}_{\mathcal{G}_n}(\cdot)$, respectively. Also, for $x \in \mathbb{R}$

$$Q_n(x) = \mathbb{P}_{\mathcal{F}_n}\left(\|S_R\|_\infty > x\right), \qquad \tilde{Q}_n(x) = \mathbb{P}_{\mathcal{G}_n}(\|\tilde{S}_R\|_\infty > x), \qquad \hat{Q}_n(x) = \mathbb{P}_{\mathcal{G}_n}(\|\tilde{S}_{\hat{R}}\|_\infty > x).$$

Define $a_{n,1} = \sup_{x \in \mathbb{R}} |Q_n(x) - \tilde{Q}_n(x)|$ and $a_{n,2} = \sup_{x \in \mathbb{R}} |\tilde{Q}_n(x) - \hat{Q}_n(x)|$.

***Proof of Theorem 3.*** The proof consists of three steps. First, we show the consistency in approximating $\mathbb{P}_{\mathcal{F}_n}(\|S_R\|_\infty \leq x)$ with $\mathbb{P}_{\mathcal{G}_n}(\|\tilde{S}_R\|_\infty \leq x)$; second, we show the equivalence between $\mathbb{P}_{\mathcal{G}_n}(\|\tilde{S}_R\|_\infty \leq x)$ and $\mathbb{P}_{\mathcal{G}_n}(\|\tilde{S}_{\hat{R}}\|_\infty \leq x)$; finally, we show the desired result.

We begin by decomposing $T_n$ (4.1) as follows

$$T_n = \|\hat{S}_n\|_\infty \text{ with } \hat{S}_n = \Delta_n + S_R, \tag{B.2}$$

where $\Delta_n := \sqrt{n}(\hat{\beta}_u - \beta_d - \hat{\delta} - S_R)$.

For this end, we define an event $\mathcal{J}_n$ as

$$\mathcal{J}_n := \left\{ \min_{1 \leq j \leq p} n^{-1} \sum_{i=1}^n \mathbb{E}(R_{i,j}^2 \mid \mathcal{F}_n) > b \text{ and } \min_{1 \leq j \leq p} n^{-1} \sum_{i=1}^n (R_{i,j} - \bar{R}_j)^2 > b/2 \right\}. \tag{B.3}$$

and observe that based on Assumption 1 (ii)-(iv), we have $\mathbb{P}(\mathcal{J}_n) \to 1$.

**Step 1: The validity of approximating $\mathbb{P}_{\mathcal{F}_n}(\|S_R\|_\infty \leq x)$ with $\mathbb{P}_{\mathcal{G}_n}(\|\tilde{S}_R\|_\infty \leq x)$.**

Let $\{\Phi_i\}_{i=1}^n$ be a sequence of random elements in $\mathbb{R}^p$ such that conditional on $\mathcal{F}_n$, $\{\Phi_i\}_{i=1}^n$ is independent across $i$ and $\Phi_i \mid \mathcal{F}_n$ is Gaussian with mean zero and variance $\mathbb{E}(R_i R_i^\top \mid \mathcal{F}_n)$. Notice that for any $x \in \mathbb{R}$, $\{a \in \mathbb{R}^p \mid \|a\|_\infty \leq x\}$ is rectangle in $\mathbb{R}^p$. By Proposition 2.1 of Chernozhukov et al. (2014) applied to the conditional probability measure $\mathbb{P}_{\mathcal{F}_n}(\cdot)$, we have on the event $\mathcal{J}_n$

$$\sup_{x \in \mathbb{R}} |\mathbb{P}_{\mathcal{F}_n}\left(\|S_R\|_\infty \leq x\right) - \mathbb{P}_{\mathcal{F}_n}\left(\|S_\Phi\|_\infty \leq x\right)| \leq C_1 D_n \quad a.s, \tag{B.4}$$

where $C_1 > 0$ is a constant depending only on $b$ and $D_n = (n^{-1} B_n^2 \log^7(pn))^{1/6}$.

Applying Corollary 4.2 of Chernozhukov et al. (2014) to the conditional probability measure $\mathbb{P}_{\mathcal{F}_n}(\cdot)$, on the event $\mathcal{J}_n$, we obtain that, for $\alpha_n = \exp\{-[(n^{-1}B_n \log^5(pn))^{-1/4} \vee 1]\}$,

$$\mathbb{P}_{\mathcal{F}_n}\left[\sup_{x \in \mathbb{R}} \left|\mathbb{P}_{\mathcal{F}_n}(\|S_\Phi\|_\infty \leq x) - \mathbb{P}_{\mathcal{G}_n}\left(\|\tilde{S}_R\|_\infty \leq x\right)\right| > C_2 \tilde{D}_n\right] \leq \alpha_n \quad a.s, \tag{B.5}$$

where $C_2 > 0$ is a constant depending only on $b$ and $\tilde{D}_n = (n^{-1}B_n^2 \log^5(pn) \log^2(\alpha_n^{-1}))^{1/6}$. Straightforward computations show that $\alpha_n$, $\tilde{D}_n$ and $D_n$ are $o_P(1)$. By (B.3), (B.4) and (B.5), we have

$$\sup_{x \in \mathbb{R}} \left|\mathbb{P}_{\mathcal{F}_n}(\|S_R\|_\infty \leq x) - \mathbb{P}_{\mathcal{G}_n}\left(\|\tilde{S}_R\|_\infty \leq x\right)\right| = o_P(1). \tag{B.6}$$

**Step 2: The equivalence between $\mathbb{P}_{\mathcal{G}_n}(\|\tilde{S}_R\|_\infty \leq x)$ and $\mathbb{P}_{\mathcal{G}_n}(\|\tilde{S}_{\hat{R}}\|_\infty \leq x)$.**

Define $\varepsilon_n = q_n^{1/4}$ with $q_n = \max_{1 \leq j \leq p} n^{-1} \sum_{i=1}^n (\hat{R}_{i,j} - R_i)^2$. By Lemma 16,

$$\begin{aligned}
&\sup_{x \in \mathbb{R}} \left|\mathbb{P}_{\mathcal{G}_n}\left(\|\tilde{S}_R\|_\infty > x\right) - \mathbb{P}_{\mathcal{G}_n}\left(\|\tilde{S}_{\hat{R}}\|_\infty > x\right)\right| \\
&\leq \mathbb{P}_{\mathcal{G}_n}\left(\|\tilde{S}_R - \tilde{S}_{\hat{R}}\|_\infty > \varepsilon_n\right) + \sup_{x \in \mathbb{R}} \mathbb{P}_{\mathcal{G}_n}\left(\|\tilde{S}_R\|_\infty \in (x - \varepsilon_n, x + \varepsilon_n]\right).
\end{aligned} \tag{B.7}$$

Notice that conditional on $\mathcal{G}_n$, $\tilde{S}_R$ is a zero-mean Gaussian vector whose $j$th entry has variance of $n^{-1} \sum_{i=1}^n (R_{i,j} - \bar{R}_j)^2$. Hence, by Lemma 18, there exists a constant $C_b > 0$ depending only on $b$ such that on the event $\mathcal{J}_n$

$$\sup_{x \in \mathbb{R}} \mathbb{P}_{\mathcal{G}_n}\left(\|\tilde{S}_R\|_\infty \in (x - \varepsilon_n, x + \varepsilon_n]\right) \leq C_b \varepsilon_n \sqrt{\log p}. \tag{B.8}$$

Also notice that conditional on $\mathcal{G}_n$, $\tilde{S}_R - \tilde{S}_{\hat{R}}$ is a zero-mean Gaussian vector whose $j$th entry has variance equal to

$$n^{-1} \sum_{i=1}^n \left[(\hat{R}_{i,j} - \bar{\hat{R}}_j) - (R_{i,j} - \bar{R}_j)\right]^2 = n^{-1} \sum_{i=1}^n (\hat{R}_{i,j} - R_{i,j})^2 - (\bar{\hat{R}}_j - \bar{R}_j)^2 \leq n^{-1} \sum_{i=1}^n (\hat{R}_{i,j} - R_{i,j})^2.$$

Recall that for any Gaussian random variable $Z \sim N(0, \sigma^2)$ and $x > 0$, $\mathbb{P}(|Z| > x) \leq C \exp(-C\sigma^{-2}x^2)$ for some universal constant $C > 0$. This elementary fact implies that

$$\mathbb{P}_{\mathcal{G}_n}(\|\tilde{S}_R - \tilde{S}_{\hat{R}}\|_\infty > \varepsilon_n) \leq \sum_{j=1}^p \mathbb{P}_{\mathcal{G}_n}(|\tilde{S}_{R,j} - \tilde{S}_{\hat{R},j}| > \varepsilon_n) \leq pC \exp(-C\varepsilon_n^2 q_n^{-1}). \tag{B.9}$$

Combining (B.7), (B.8) and (B.9), we have on the event $\mathcal{J}_n$

$$\sup_{x\in\mathbb{R}} \left| \mathbb{P}_{\mathcal{G}_n}\left(\|\tilde{S}_R\|_\infty > x\right) - \mathbb{P}_{\mathcal{G}_n}\left(\|\tilde{S}_{\hat{R}}\|_\infty > x\right) \right| \leq C_b \varepsilon_n \sqrt{\log p} + pC \exp(-C\varepsilon_n^2 \sigma_{n,*}^{-2}).$$

Since $\varepsilon_n^2 q_n^{-1}/\log p = (q_n \log^2 p)^{-1/2} \to \infty$, we have $p\exp(-C\varepsilon_n^2 \sigma_{n,*}^{-2}) = o(1)$. Notice that $\varepsilon_n \sqrt{\log p} = (q_n \log^2 p)^{1/4} = o(1)$ and $\mathbf{1}\{\mathcal{J}_n^c\} = o_P(1)$ (by (B.3)). The above display implies

$$\sup_{x\in\mathbb{R}} \left| \mathbb{P}_{\mathcal{G}_n}\left(\|\tilde{S}_R\|_\infty \leq x\right) - \mathbb{P}_{\mathcal{G}_n}\left(\|\tilde{S}_{\hat{R}}\|_\infty \leq x\right) \right| = o_P(1). \tag{B.10}$$

**Step 3: Approximating the null distribution**

Define $t_n = \lambda_{1,n} \vee \lambda_{2,n}$ and $s_n = t_n^{1/2} n^{1/4} \log^{-1/4} p$. Notice that, $\forall x \in \mathbb{R}$,

$$\begin{aligned}
&\left| \mathbb{P}_{\mathcal{F}_n}\left(\|S_R\|_\infty \in (x-s_n, x+s_n]\right) - \mathbb{P}_{\mathcal{G}_n}\left(\|\tilde{S}_R\|_\infty \in (x-s_n, x+s_n]\right)\right| \\
&= \left| [Q_n(x-s_n) - Q_n(x+s_n)] - [\tilde{Q}_n(x-s_n) - \tilde{Q}_n(x+s_n)] \right| \\
&\leq \left| Q_n(x-s_n) - \tilde{Q}_n(x-s_n) \right| + \left| Q_n(x+s_n) - \tilde{Q}_n(x+s_n) \right| \leq 2a_{n,1}.
\end{aligned} \tag{B.11}$$

Recall $\hat{S}_n$ and $\Delta_n$ from (B.2). By assumption, under $H_0$, $\|\Delta_n\|_\infty = O_P(\sqrt{n} t_n)$. Notice that

$$\begin{aligned}
\left| \mathbb{P}_{\mathcal{F}_n}\left(\|\hat{S}_n\|_\infty > x\right) - \tilde{Q}_n(x) \right| &\leq \left| \mathbb{P}_{\mathcal{F}_n}(\|\hat{S}_n\|_\infty > x) - Q_n(x) \right| + a_{n,1} \\
&\overset{(i)}{\leq} \mathbb{P}_{\mathcal{F}_n}(\|\Delta_n\|_\infty > s_n) + \mathbb{P}_{\mathcal{F}_n}\left(\|S_R\|_\infty \in (x-s_n, x+s_n]\right) + a_{n,1} \\
&\overset{(ii)}{\leq} \mathbb{P}_{\mathcal{F}_n}(\|\Delta_n\|_\infty > s_n) + \mathbb{P}_{\mathcal{G}_n}\left(\|\tilde{S}_R\|_\infty \in (x-s_n, x+s_n]\right) + 3a_{n,1},
\end{aligned} \tag{B.12}$$

where $(i)$ follows by Lemma 16 and $(ii)$ follows by (B.11). Notice that conditional on $\mathcal{G}_n$, $\tilde{S}_R$ is a zero-mean Gaussian vector in $\mathbb{R}^p$ whose $j$th component has variance equal to $n^{-1}\sum_{i=1}^n (R_{i,j} - \bar{R}_j)^2$. By Lemma 18, there exists a constant $C_b > 0$ depending only on $b$ such that on the event $\mathcal{J}_n$

$$\sup_{x\in\mathbb{R}} \mathbb{P}_{\mathcal{G}_n}\left(\|\tilde{S}_R\|_\infty \in (x-s_n, x+s_n]\right) \leq s_n C_b \sqrt{\log p} \quad a.s. \tag{B.13}$$

Therefore, on $\mathcal{J}_n$

$$\begin{aligned}
\sup_{x\in\mathbb{R}} \left| \mathbb{P}_{\mathcal{F}_n}\left(\|\hat{S}_n\|_\infty > x\right) - \hat{Q}_n(x) \right| &\leq \sup_{x\in\mathbb{R}} \left| \mathbb{P}_{\mathcal{F}_n}(\|\hat{S}_n\|_\infty > x) - \tilde{Q}_n(x) \right| + \sup_{x\in\mathbb{R}} |\tilde{Q}_n(x) - \tilde{Q}_n(x)| \\
&\overset{(i)}{\leq} \mathbb{P}_{\mathcal{F}_n}(\|\Delta_n\|_\infty > s_n) + s_n C_b \sqrt{\log p} + 3a_{n,1} + a_{n,2}
\end{aligned}$$

$$\stackrel{(ii)}{=} o_P(1), \tag{B.14}$$

where $(i)$ follows by (B.12), (B.13) and the definition of $a_{n,2}$, whereas $(ii)$ follows by $\mathbb{P}(\mathcal{J}_n^c) = o(1)$, $a_{n,1} = o_P(1)$ (by (B.6)) and $a_{n,2} = o_P(1)$ (by (B.10)), together with $\|\Delta_n\|_\infty/s_n = O_P(\sqrt{n}t_n/s_n) = O_P((t_n^2 n \log p)^{-1/4}) = o_P(1)$ and $s_n\sqrt{\log p} = (t_n^2 n \log p)^{1/4} = o(1)$. It follows that $\forall \delta > 0$,

$$
\begin{aligned}
&\mathbb{E}\left[\sup_{\eta \in (0,1)} \left|\mathbb{P}_{\mathcal{F}_n}\left(\|\hat{S}_n\|_\infty > \mathcal{Q}(1-\eta, \|\tilde{S}_{\hat{R}}\|_\infty)\right) - \eta\right|\right] \\
&= \mathbb{E}\left[\sup_{x \in \mathbb{R}} \left|\mathbb{P}_{\mathcal{F}_n}\left(\|\hat{S}_n\|_\infty > \hat{Q}_n^{-1}(\eta)\right) - \eta\right|\right] \\
&\stackrel{(i)}{\leq} \mathbb{E}\left[\delta + \mathbb{P}_{\mathcal{F}_n}\left(\sup_{x \in \mathbb{R}} \left|\mathbb{P}_{\mathcal{F}_n}(\|\hat{S}_n\|_\infty > x) - \hat{Q}_n(x)\right| > \delta\right)\right] \\
&= \delta + \mathbb{P}\left(\sup_{x \in \mathbb{R}} \left|\mathbb{P}_{\mathcal{F}_n}(\|\hat{S}_n\|_\infty > x) - \hat{Q}_n(x)\right| > \delta\right) \stackrel{(ii)}{\leq} \delta + o(1)
\end{aligned}
\tag{B.15}
$$

where $(i)$ follows by Lemma 17 and $(ii)$ follows by (B.14). Since $\delta$ is arbitrary, (B.15) implies

$$\mathbb{E}\left[\sup_{\eta \in (0,1)} \left|\mathbb{P}_{\mathcal{F}_n}\left(\|\hat{S}_n\|_\infty > \mathcal{Q}(1-\eta, \|\tilde{S}_{\hat{R}}\|_\infty)\right) - \eta\right|\right] = o(1).$$

The desired result follows by noticing that $\sup_\eta |\mathbb{E}(Z_n(\eta))| \leq \mathbb{E}\sup_\eta |Z_n(\eta)|$, where $Z_n(\eta) = \mathbb{P}_{\mathcal{F}_n}\left(\|\hat{S}_n\|_\infty > \mathcal{Q}(1-\eta, \|\tilde{S}_{\hat{R}}\|_\infty)\right) - \eta$. $\square$

### B.2 Proof of Theorem 4

**Proof of Theorem 4.** We first derive the rates for the critical value. Let $\hat{\Omega} = n^{-1}\sum_{i=1}^n \hat{R}_i \hat{R}_i^\top$ and $\Omega = n^{-1}\sum_{i=1}^n R_i R_i^\top$. Recall $\tilde{S}_R$ and $\tilde{S}_{\hat{R}}$ defined in the proof of Theorem 3. Notice that conditional on $\mathcal{G}_n$, $\tilde{S}_{\hat{R}} \sim N(0, \hat{\Omega})$.

Recall that for any Gaussian random variable $Z \sim N(0, \sigma^2)$ and $x > 0$, $\mathbb{P}(|Z| > x) \leq C\exp(-C\sigma^{-2}x^2)$ for some universal constant $C > 0$. The union bound implies that $P(\|\tilde{S}_{\hat{R}}\|_\infty > t \mid \mathcal{G}_n) \leq 2p\exp(-Ct^2/\max_{1\leq j\leq p}\hat{\Omega}_{j,j})$. Since $\max_{1\leq j\leq p}\sum_{i=1}^n(\hat{R}_{i,j} - R_{i,j})^2 = o_P(n/\log^2(p \vee n))$, we have that $\|\hat{\Omega} - \Omega\|_\infty = o_P(\log^{-2}(p \vee n)) = o_P(1)$. Since $\max_{1\leq j\leq p}\Omega_{j,j} = O(1)$, we have that for any $\alpha \in (0,1)$,

$$\mathcal{Q}(1-\alpha, T_n^{BS}) = O_P(\sqrt{\log p}). \tag{B.16}$$

By triangular inequality, we have

$$
\begin{aligned}
T_n &= \sqrt{n}\|\hat{\beta}_u - \hat{\beta}_d - \hat{\delta}\|_\infty \\
&= \sqrt{n}\|(\hat{\beta}_u - \beta_*) - (\hat{\beta}_d - \beta_*) - \hat{\delta}\|_\infty \\
&\geq \sqrt{n}\|\hat{\beta}_d - \beta_*\|_\infty - \sqrt{n}\|\hat{\beta}_u - \beta_*\|_\infty - \sqrt{n}\|\hat{\delta}\|_\infty \\
&\overset{(i)}{\geq} \sqrt{n}c_n - \sqrt{n}\|\hat{\beta}_u - \beta_*\|_1 - \sqrt{n}\|\hat{\delta}\|_\infty \\
&\overset{(ii)}{=} \sqrt{n}c_n - O_P(n^{-1/4}),
\end{aligned}
\tag{B.17}
$$

where $(i)$ follows by the fact that $\hat{\beta}_d \in \mathcal{B}_0$ and thus $\|\hat{\beta}_d - \beta_*\|_\infty \geq \min_{\beta \in \mathcal{B}_0} \|\beta - \beta_*\|_\infty \geq c_n$ and $(ii)$ follows by the definition of $\hat{\delta}$ and Assumption 2.

Since $c_n \gg n^{-1/4} \gg \sqrt{n^{-1}\log p}$, the desired result follows by (B.16) and (B.17). □

## C  Proof of results for Linear Model and Logistic Model

The following result is useful in proving both theorems in Section 5.

**Lemma 6.** *Let $X$ be a random variable. Suppose that there exists a constant $c > 0$ such that $\mathbb{P}(|X| > t) \leq \exp(1 - ct^2)$ for all $t > 0$. Then, $\mathbb{E}\exp(|X|/D) < 2$, where $D \geq \sqrt{7/[c\log(3/2)]}$.*

### C.1  Proof of Theorems 1 and 2

We begin by providing five useful lemmas for the proof. Lemma 7 provides an upper bounded for the estimation error of the initial estimator $\hat{\beta}_d$ as well as its feasibility.

**Lemma 7.** *Consider model (3.1) and the estimator $\hat{\beta}_d$ (2.1). Let Assumptions of Theorem 1 hold. Then the constraint set of the optimization problem (2.1) is non-empty with probability approaching one, $\|\hat{\beta}_u - \beta_*\|_1 = O_P(\|\beta_*\|_0\sqrt{n^{-1}\log p})$ and $\|X(\hat{\beta}_u - \beta_*)\|_2^2 = O_P(\|\beta_*\|_0 \log p)$.*

The above lemma is the extension of the results derived for the Dantzig selector (Theorem 7.1 of Bickel et al. (2009)) to the designs that are have sub-Gaussian tails. It only suffices to show that the design matrix $X$ satisfies the restricted eigenvalue condition, which is guaranteed by Theorem 6 of Rudelson and Zhou (2013). We omit the details for brevity.

Now in order to prove the theorem, let us define two quantities and treat them separately in the following two lemmas. Let $\mathcal{F}_n$ denote the $\sigma$-algebra generated by the design matrix $X$. Recall that for the linear regression model (3.1), we have $s(W_i, \beta) = -x_i(y_i - x_i^\top \beta)$ and

$$R_i = \begin{cases} -2\hat{\Theta} x_i \varepsilon_i & 1 \leq i \leq m \\ 2\hat{\Theta} x_i \varepsilon_i & m+1 \leq i \leq n. \end{cases} \quad (C.1)$$

Moreover, let $\Omega_X = \Sigma_X^{-1}$. For $1 \leq j \leq p$, we denote by $\Omega_{X,j}$ the $j$th column of $\Omega_X$.

**Lemma 8.** *Consider model (3.1) and suppose that the assumptions of Theorem 1 hold. Then there exist constants $a_1, a_2 > 0$ depending only on $\kappa_1$, $\kappa_2$ and $\kappa_3$ such that if we choose $\eta \geq a_1\sqrt{n^{-1}\log p}$ and $\mu \geq a_2\sqrt{\log(p \vee n)}$ in (3.4), then $\mathbb{P}(\mathcal{A}) \to 1$, where*

$$\mathcal{A} = \bigcap_{j=1}^{p} \{\Omega_{X,j} \text{ is feasible in (3.4)}\}. \quad (C.2)$$

**Lemma 9.** *Consider model (3.1) and $\hat{\beta}_u$ as in (2.1). Let the assumptions of Theorem 1 hold. Then under $H_0$ in (1.2),*

$$\max_{1 \leq j \leq p} n^{-1} \sum_{i=1}^{n} (\hat{R}_{i,j} - R_{i,j})^2 = O_P\left(n^{-1}\|\beta_*\|_0 \log^2(p \vee n)\right),$$

*where $\hat{R}_i$ and $R_i$ are defined in (5.8) and (C.1), respectively.*

The next two lemmas discuss validity of the various parts of the Assumption 1 needed for the successful approximation of the null distribution. The first verifies "Lyapunov-type" conditions whereas the second verifies the assumptions of the bootstrap approximation.

**Lemma 10.** *Consider model (3.1) and let the assumptions of Theorem 1 hold. Let $R_i$ be defined in (C.1). Then there exists a constant $K > 0$ such that for*

$$B_n = 8K \left(\|X\hat{\Theta}\|_\infty \vee 1\right)^3,$$

*we have that, almost surely, (1) $\max_{1 \leq i \leq n, 1 \leq j \leq p} \mathbb{E}[\exp(|R_{i,j}|/B_n) \mid \mathcal{F}_n] \leq 2$; (2) $\max_{1 \leq j \leq p} \sum_{i=1}^{n} \mathbb{E}(|R_{i,j}|^3 \mid \mathcal{F}_n) \leq nB_n$; (3) $\max_{1 \leq j \leq p} \sum_{i=1}^{n} \mathbb{E}(R_{i,j}^4 \mid \mathcal{F}_n) \leq nB_n^2$.*

**Lemma 11.** *Consider model (3.1) and let Assumptions of Theorem 1 hold. Let $R_i$ be defined in (C.1). Then,*

(1) *there exist a constant $b > 0$ such that $\mathbb{P}\left(\min_{1 \leq j \leq p} n^{-1} \sum_{i=1}^{n} \mathbb{E}(R_{i,j}^2 \mid \mathcal{F}_n) \geq b\right) \to 1$,*

(2) $\|n^{-1} \sum_{i=1}^n R_i\|_\infty = o_P(1)$,

(3) $\max_{1 \leq j \leq p} \left| n^{-1} \sum_{i=1}^n [R_{i,j}^2 - \mathbb{E}(R_{i,j}^2 \mid \mathcal{F}_n)] \right| = o_P(1)$.

The proofs of the above five lemmas will be given in Appendix D.

**Proof of Theorem 1.** By Theorem 3, it suffices to verify Assumption 1 (i)-(v), Assumption 2 and $\max_{1 \leq j \leq p} \sum_{i=1}^n (\hat{R}_{i,j} - R_{i,j})^2 = o_P(n/\log^2(p \vee n))$.

Observe that for the linear model $l(z_i, \beta) = (y_i - x_i^\top \beta)^2/2$ with $z_i = (y_i, x_i)$ leading to $s(z_i, \beta) = x_i(x_i^\top \beta - y_i)$ and $\nabla_\beta^2 l(z_i, \beta) = x_i x_i^\top$.

(1) We first show Assumption 2. Under the notation of Assumption 2, $\hat{H}_A(\beta) = m^{-1} X_A^\top X_A$ and $\hat{H}_B = m^{-1} X_B^\top X_B$. Notice that

$$\sup_{t \in [0,1]} \|[\mathbb{I}_p - \hat{\Theta}_A \hat{H}_A(\hat{\beta}_u - t(\hat{\beta}_u - \beta_*))](\hat{\beta}_u - \beta_*)\|_\infty$$
$$= \|[\mathbb{I}_p - \hat{\Theta} X_A^\top X_A/m](\hat{\beta}_u - \beta_*)\|_\infty$$
$$\overset{(i)}{\leq} \|\mathbb{I}_p - \hat{\Theta} X_A^\top X_A/m\|_\infty \|\hat{\beta}_u - \beta_*\|_1$$
$$\overset{(ii)}{=} O_P(\|\beta_*\|_0 n^{-1} \log p),$$

where $(i)$ follows by Holder's inequality and $(ii)$ follows by $\|\mathbb{I}_p - \hat{\Theta} X_A^\top X_A/m\|_\infty = O_P(\sqrt{n^{-1}\log p})$ (the constraint in (3.4) and Lemma 8) and $\|\hat{\beta}_u - \beta_*\|_1 = O_P(\|\beta_*\|_0 \sqrt{n^{-1}\log p})$ (Lemma 7). Similarly, we have that, under $H_0$,

$$\sup_{t \in [0,1]} \|[\mathbb{I}_p - \hat{\Theta}_B \hat{H}_B(\hat{\beta}_d - t(\hat{\beta}_d - \beta_*))](\hat{\beta}_d - \beta_*)\|_\infty$$
$$\leq \|\mathbb{I}_p - \hat{\Theta} X_B^\top X_B/m\|_\infty \|\hat{\beta}_d - \beta_*\|_1 \overset{(i)}{=} O_P(\|\beta_*\|_0 n^{-1} \log p),$$

where $(i)$ follows by $\|\mathbb{I}_p - \hat{\Theta} X_B^\top X_B/m\|_\infty = O_P(\sqrt{n^{-1}\log p})$ (the constraint in (3.4) and Lemma 8) and $\|\hat{\beta}_d - \beta_*\|_1 = O_P(\|\beta_*\|_0 \sqrt{n^{-1}\log p})$ (Lemma 7 and Lemma 1). By the above two displays and the rate for $\|\beta_*\|_0$, Assumption 2(i) holds.

Since $\hat{\Theta}$ is $\mathcal{F}_n$-measurable, we can take $\Theta_A = \Theta_B = \hat{\Theta}$. Thus, Assumption 2(ii) holds. Lastly, Lemma 7 and $\|\beta_*\|_0 = o(n^{1/4}/\sqrt{\log p})$ imply $\|\hat{\beta}_u - \beta_*\|_1 = o(n^{-1/4})$. Assumption 2 holds.

(2) Since $\varepsilon$ is independent of $X$ and $\hat{\Theta}$ only depends on $X$, Assumption 1 (i) holds.

(3) Assumption 1 (ii) and (iii) hold by Lemma 11.

(4) By Lemma 10, the first three inequalities in Assumption 1 (v) hold with $B_n = O(\|X\hat{\Theta}\|_\infty^3 \vee 1)$. Notice that on the event $\mathcal{A}$ defined in (C.2), $\|X\hat{\Theta}\|_\infty \leq \mu = O(\sqrt{\log(p \vee n)})$. Since $\mathbb{P}(\mathcal{A}) \to 1$ (Lemma 8), $B_n = O_P(\log^{3/2}(p \vee n))$. Simple computations yield the last inequality in Assumption 1(iv).

(5) By Lemma 9, $\max_{1 \leq j \leq p} \sum_{i=1}^n (\hat{R}_{i,j} - R_{i,j})^2 = O_P(\|\beta_*\|_0 \log^2(p \vee n)) \stackrel{(i)}{=} o_P(n/\log^2(p \vee n))$, where $(i)$ holds by $\|\beta_*\|_0 = o(n^{1/4}/\sqrt{\log p})$ and $\log p = o(n^{1/8})$.

The proof is complete. □

***Proof of Theorem 2***. We apply Theorem 4. Notice that the conditions we need to verify are exactly the same as in the proof of Theorem 1. Hence, the same arguments apply. □

## C.2 Proof of Theorems 5 and 6

We begin by providing four lemmas, tailored to the logistic regression model, which are useful for the proof. We define $s_\Theta = \max_{1 \leq j \leq p} \|\Theta_{*,j}\|_0$ and $s_\beta = \|\beta_*\|_0$.

**Lemma 12.** *Consider the model (5.2) and $\hat{\beta}_u$ in (5.3). Let assumptions of Theorem 5 hold. Then $\|\hat{\beta}_u - \beta_*\|_1 = O_P(s_\beta \sqrt{n^{-1} \log p})$.*

The above lemma is the extension of the results derived for the Lasso estimator in the generalized linear models (see proof of Theorem 3.3 Van de Geer et al. (2014) for condition (C2) therein) to the designs that have sub-Gaussian tails. With similar techniques to those developed in Van de Geer (2008), we can obtain the above lemma. The details are omitted for the brevity.

Throughout this subsection, we use the definitions in Section 5. We introduce the following notations:

$$R_i = \begin{cases} -2\Theta_* x_i [-y_i + b'(x_i^\top \beta_*)] & 1 \leq i \leq m \\ 2\Theta_* x_i [-y_i + b'(x_i^\top \beta_*)] & m+1 \leq i \leq n. \end{cases} \quad (C.3)$$

For for $z_i = (y_i, x_i)$, let $s(z, \beta) = x_i[-y_i + b'(x_i^\top \beta)]$ with $\hat{H}_A(\beta) = m^{-1} \sum_{i=1}^m x_i x_i^\top b''(x_i^\top \beta)$ and $\hat{H}_B(\beta) = m^{-1} \sum_{i=m+1}^n x_i x_i^\top b''(x_i^\top \beta)$.

The next two lemmas verify Assumption 2 for the case of logistic regression model.

**Lemma 13.** *Consider the model (5.2) and $\hat{\beta}_u$ in (5.3). Let assumptions of Theorem 5 hold. Then*

$$\sup_{t\in[0,1]} \left\|[\mathbb{I}_p - \hat{\Theta}_A \hat{H}_A(\hat{\beta}_u - t(\hat{\beta}_u - \beta_*))](\hat{\beta}_u - \beta_*)\right\|_\infty \vee \left\|(\hat{\Theta}_A - \Theta_*)m^{-1}\sum_{i=1}^m s(z_i, \beta_*)\right\|_\infty$$

$$= O_P\left(\left[s_\Theta \vee \left(s_\beta^2 \log^{3/2}(p \vee n)\right)\right]n^{-1}\log^{3/2}(p \vee n)\right).$$

**Lemma 14.** *Consider the model (5.2) and $\hat{\beta}_u$ in (5.3). Let assumptions of Theorem 5 hold. Then,*

$$\sup_{t\in[0,1]} \left\|[\mathbb{I}_p - \hat{\Theta}_B \hat{H}_B(\hat{\beta}_d - t(\hat{\beta}_d - \beta_*))](\hat{\beta}_d - \beta_*)\right\|_\infty \vee \left\|(\hat{\Theta}_B - \Theta_*)m^{-1}\sum_{i=1}^m s(z_i, \beta_*)\right\|_\infty$$

$$= O_P\left(\left[s_\Theta \vee \left(s_\beta^2 \log^{3/2}(p \vee n)\right)\right]n^{-1}\log^{3/2}(p \vee n)\right).$$

Lastly, we present Lemma on the error of the plug-in estimates $\hat{R}_i$s in estimating the main term of the linearization step of the Algorithm 1.

**Lemma 15.** *Consider the model (5.2), $\hat{\beta}_u$ in (5.3), $R_i$ in (C.3) and $\hat{R}_i$ of Section 5. Let assumptions of Theorem 5 hold. Then*

$$\max_{1\leq j\leq p} n^{-1}\sum_{i=1}^n (\hat{R}_{i,j} - R_{i,j})^2 = O_P([s_\Theta^2 \vee (K_n^2 s_\beta^2)]K_n^4 n^{-1}\log p),$$

*where $K_n = \sqrt{\log(p \vee n)}$.*

The proofs of the above four lemmas will be given in Appendix D.

***Proof of Theorem 5.*** Consider $R_i$ defined in (C.3). By Theorem 3, it suffices to verify Assumption 1 (i)-(v), Assumption 2 and $\max_{1\leq j\leq p}\sum_{i=1}^n(\hat{R}_{i,j} - R_{i,j})^2 = o_P(n/\log^2(p \vee n))$, where $\hat{R}_i$ is defined in (5.8).

(1) By Lemmas 13 and 14, the rates $\lambda_{1,n}, \lambda_{2,n}$ in Assumption 2(i)-(ii) are $\lambda_{1,n} = \lambda_{2,n} = \left[s_\Theta \vee \left(s_\beta^2 \log^3(p \vee n)\right)\right]n^{-1}\log^{3/2}(p \vee n)$, where $\mathcal{F}_n$ is the $\sigma$-algebra generated by $X$. Simple computation yields $\lambda_{1,n} \vee \lambda_{2,n} = o(\sqrt{n^{-1}\log p})$. Lastly, by Lemma 12 and $s_\beta = o([n/\log^5 p]^{1/4})$, we have $\|\hat{\beta}_u - \beta_*\|_1 = o(n^{-1/4})$. Hence, Assumption 2 holds.

(2) By (5.1), $\mathbb{E}[-y_i + b'(x_i^\top \beta_*) \mid X] = \mathbb{E}[\exp(x_i^\top \beta_*)/(1 + \exp(x_i^\top \beta_*)) - y_i \mid X] = 0$. Thus, $\mathbb{E}R_i = 0$ and Assumption 1 (i) holds.

(3) To verify Assumption 1 (ii) and (iii) notice that $|R_{i,j}| = 2|\Theta_{*,j}^\top x_i| \cdot |b'(x'\beta_*) - y_i|$. Also notice that $\Theta_{*,j}^\top x_i$ has bounded sub-Gaussian norm and both $|b'(x_i'\beta_*)|$ and $|y_i|$ are bounded by 1. Therefore, $R_{i,j}$ has bounded sub-Gaussian norms and $R_{i,j}^2$ has bounded sub-exponential norms. By Hoeffding's inequality and Bernstein's inequality, together with the union bound, we have that
$$\begin{cases} \max_{1\leq j\leq p} |n^{-1} \sum_{i=1}^n R_{i,j} - \mathbb{E} R_{i,j}| = o_P(1) \\ \max_{1\leq j\leq p} |n^{-1} \sum_{i=1}^n [R_{i,j}^2 - \mathbb{E} R_{i,j}^2]| = o_P(1). \end{cases}$$

Assumption 1 (ii) follows by $\mathbb{E} R_{i,j} = 0$. Notice that

$$\min_{1\leq j\leq p} n^{-1} \sum_{i=1}^n \mathbb{E}(R_{i,j}^2) = 4 \min_{1\leq j\leq p} \Theta_{*,j}^\top \left( \mathbb{E} x_1 x_1^\top [-y_1 + b'(x_1^\top \beta_*)]^2 \right) \Theta_{*,j} = 4 \min_{1\leq j\leq p} \Theta_{*,j}^\top \Theta_*^{-1} \Theta_{*,j}$$

is bounded away from zero by assumptions of Theorem 5. Assumption 1 (iii) follows.

(4) Finally, we show Assumption 1(iv). Since $R_{i,j}$ has bounded sub-Gaussian norms, it follows, by Lemma 6, that there exists a constant $D_1 > 0$ such that $\mathbb{E} \exp(|R_{i,j}|/D_1) \leq 2$. The sub-Gaussian property also implies that there exists a constant $D_2 > 0$ such that $\mathbb{E}|R_{i,j}|^3 \leq D_2$ and $\mathbb{E} R_{i,j}^4 \leq D_2$. Hence, Assumption 1 (iv) holds with $B_n = D_2 \vee \sqrt{D_2} \vee D1$.

(5) By Lemma 15 and the rate conditions for $s_\Theta$ and $s_\beta$, we have $\max_{1\leq j\leq p} \sum_{i=1}^n (\hat{R}_{i,j} - R_{i,j})^2 = o_P(n/\log^2(p \vee n))$.

The proof is complete. □

**Proof of Theorem 6.** We apply Theorem 4. Notice that the conditions we need to verify are exactly the same as in the proof of Theorem 5. Hence, the same arguments apply. □

## D  Proofs of Lemmas

*Proof of Lemma 6.* Let $Z = \exp(|X|/D)$. Since $Z \geq 1$, we have the decomposition $Z = \sum_{i=1}^\infty Z \mathbf{1}\{i - 1/2 < Z \leq i + 1/2\}$. Define the sequence $b_i = (i - 1/2)^2 \exp[-cD^2 \log^2(i - 1/2)]$. By Fubini's theorem,

$$\mathbb{E} Z = \sum_{i=1}^\infty \mathbb{E} Z \mathbf{1}\{i - 1/2 < Z \leq i + 1/2\} \leq \sum_{i=1}^\infty (i + 1/2) \mathbb{P}(Z > i - 1/2)$$
$$= 3/2 + \sum_{i=2}^\infty (i + 1/2) \mathbb{P}(Z > i - 1/2)$$

$$= 3/2 + \sum_{i=2}^{\infty}(i+1/2)\mathbb{P}\left[|X| > D\log(i-1/2)\right]$$

$$\overset{(i)}{\leq} 3/2 + e\sum_{i=2}^{\infty}(i+1/2)\exp[-cD^2\log^2(i-1/2)],$$

$$\overset{(ii)}{\leq} 3/2 + e\sum_{i=2}^{\infty} b_i, \tag{D.1}$$

where $(i)$ follows by $\mathbb{P}(|X| > t) \leq \exp(1 - ct^2) \ \forall t > 0$ and $(ii)$ follows by the elementary inequality that $(i + 1/2) \leq (i - 1/2)^2$ for $i \geq 2$. Notice that, for $i \geq 2$,

$$\log b_i - \log(i^{-4}) \leq \log b_i - \log(i-1/2)^{-4} = \left[6 - cD^2 \log(i-1/2)\right]\log(i-1/2)$$

$$\overset{(i)}{\leq} \left[6 - cD^2 \log(3/2)\right]\log(i-1/2)$$

$$\overset{(ii)}{\leq} -\log(i-1/2) \leq 0,$$

where $(i)$ holds by $i \geq 2$ and $(ii)$ holds by the definition of $D$ in the statement of the lemma. The above display implies that, $\forall i \geq 2$, $b_i \leq i^{-4}$. It follows, by (D.1), that

$$\mathbb{E}Z \leq 3/2 + e\sum_{i=2}^{\infty} i^{-4}.$$

It can be shown that $\sum_{i=2}^{\infty} i^{-4} = \pi^4/90 - 1 \leq 1/10$. Thus, $\mathbb{E}Z < 3/2 + e/10 < 2$. □

*Proof of Lemma 8.* Notice that for $1 \leq j \leq p$,

$$m^{-1}X_A^\top X_A \Omega_{X,j} - e_j = m^{-1}\sum_{i=1}^{m}(x_i x_i^\top \Omega_{X,j} - e_j).$$

Since $x_i^\top \Omega_{X,j}$ has bounded sub-Gaussian norm, it follows, by Lemma 19, that entries of $x_i x_i^\top \Omega_{X,j}$ have sub-exponential norms upper bounded by a constant $K_1 > 0$. Notice that $\mathbb{E}(x_i x_i^\top \Omega_{X,j} - e_j) = 0$. By Proposition 5.16 in Vershynin (2010) and the union bound, that for any $t > 0$,

$$\mathbb{P}\left(\max_{1 \leq j \leq p}\left\|m^{-1}\sum_{i=1}^{m}(x_i x_i^\top \Omega_{X,j} - e_j)\right\|_\infty > t\right)$$

$$\leq \sum_{j=1}^{p}\sum_{k=1}^{p}\mathbb{P}\left(\left|m^{-1}\sum_{i=1}^{m}(x_{i,k}x_i^\top \Omega_{X,j} - \mathbf{1}\{k=j\})\right| > t\right) \leq 2p^2 \exp\left[-c\min\left(\frac{mt^2}{K_1^2}, \frac{mt}{K_1}\right)\right],$$

where $c > 0$ is a universal constant. Therefore,

$$\mathbb{P}\left(\max_{1\leq j\leq p}\left\|m^{-1}\sum_{i=1}^{m}(x_i x_i^\top \Omega_{X,j} - e_j)\right\|_\infty > 4c^{-1}K_1\sqrt{m^{-1}\log p}\right) \to 0. \tag{D.2}$$

Similarly, we can show that

$$\mathbb{P}\left(\max_{1\leq j\leq p}\left\|m^{-1}\sum_{i=m+1}^{n}(x_i x_i^\top \Omega_{X,j} - e_j)\right\|_\infty > 4c^{-1}K_1\sqrt{m^{-1}\log p}\right) \to 0. \tag{D.3}$$

Let $K_2 > 0$ be a constant that upper bounds the sub-Gaussian norm of $x_i^\top \Omega_{X,j}$; such a constant exists because $\Sigma_X$ has eigenvalues bounded away from zero and infinity and $x_i$ has bounded sub-Gaussian norms. By the union bound, we have that $\forall t > 0$,

$$\mathbb{P}\left(\max_{1\leq j\leq p}\|X_A\Omega_{X,j}\|_\infty > t\right) \leq \sum_{i=1}^{n}\sum_{j=1}^{p}\mathbb{P}\left(|x_i^\top \Omega_{X,j}| > t\right) \leq np\exp\left(1 - c_0^2 t^2 K_2^{-2}\right),$$

where $c_0 > 0$ is a universal constant and the last inequality holds by the sub-Gaussian condition (see Lemma 5.5 of Vershynin (2010)). Thus,

$$\mathbb{P}\left(\max_{1\leq j\leq p}\|X_A\Omega_{X,j}\|_\infty > 4K_2 c_0^{-1}\sqrt{\log(p\vee n)}\right) \to 0. \tag{D.4}$$

Similarly, we can show that

$$\mathbb{P}\left(\max_{1\leq j\leq p}\|X_B\Omega_{X,j}\|_\infty > 4K_2 c_0^{-1}\sqrt{\log(p\vee n)}\right) \to 0. \tag{D.5}$$

The desired result follows by (D.2), (D.3), (D.4) and (D.5). $\square$

*Proof of Lemma 9.* Let $\hat{\varepsilon}_i = y_i - x_i^\top \hat{\beta}_u$. Notice that for $1 \leq j \leq p$,

$$\hat{R}_{i,j} - R_{i,j} = \begin{cases} -2\hat{\Theta}_j^\top x_i(\hat{\varepsilon}_i - \varepsilon_i) & 1 \leq i \leq m \\ 2\hat{\Theta}_j^\top x_i(\hat{\varepsilon}_i - \varepsilon_i) & m+1 \leq i \leq n. \end{cases} \tag{D.6}$$

Therefore, on the event $\mathcal{A}$,

$$\max_{1\leq j\leq p} n^{-1}\sum_{i=1}^{n}(\hat{R}_{i,j} - R_{i,j})^2 \overset{(i)}{\leq} \max_{1\leq j\leq p} n^{-1}\sum_{i=1}^{m} 4(\hat{\Theta}_j^\top x_i)^2(\hat{\varepsilon}_i - \varepsilon_i)^2$$

$$+ \max_{1\leq j\leq p} n^{-1} \sum_{i=m+1}^{n} 4(\hat{\Theta}_j^\top x_i)^2 (\hat{\varepsilon}_i - \varepsilon_i)^2$$

$$\overset{(ii)}{\leq} 4\mu^2 n^{-1} \sum_{i=1}^{m} (\hat{\varepsilon}_i - \varepsilon_i)^2 + 4\mu^2 n^{-1} \sum_{i=m+1}^{n} (\hat{\varepsilon}_i - \varepsilon_i)^2$$

$$= 4\mu^2 n^{-1} \sum_{i=1}^{n} (\hat{\varepsilon}_i - \varepsilon_i)^2 = 4\mu^2 n^{-1} \|X(\hat{\beta}_u - \beta_*)\|_2^2,$$

where $(i)$ follows by (D.6) and $(ii)$ follows by $\|X\hat{\Theta}\|_\infty \leq \mu$ on $\mathcal{A}$ (the constraint in (3.4)). The desired result follows by the above display, together with $\mathbb{P}(\mathcal{A}) \to 1$ (Lemma 8), $\|X(\hat{\beta}_u - \beta_*)\|_2^2 = O_P(\|\beta_*\|_0 \log p)$ (Lemma 7) and $\mu = O(\sqrt{\log(p \vee n)})$. □

*Proof of Lemma 10.* Since $\varepsilon_i$ has a sub-Gaussian norm bounded by a constant $K_1 > 0$, it follows by Lemma 6, that there exists a constant $K_2 > 0$ with $\mathbb{E} \exp(|\varepsilon_i|/K_2) \leq 2$. Let $K_3 > K_2$ be a constant such that $\mathbb{E}|\varepsilon_1|^3 \leq K_3$ and $\mathbb{E}|\varepsilon_1|^4 \leq K_3^2$. Define

$$B_n = 8K_3 \left( \|X\hat{\Theta}\|_\infty \vee 1 \right)^3.$$

To see part (1), notice that $\mathbb{E}[\exp(|R_{i,j}|/B_n) \mid \mathcal{F}_n] \leq \mathbb{E}[\exp(|\varepsilon_i|/K_2) \mid \mathcal{F}_n] \leq 2$. Part (2) follows by

$$\max_{1\leq j\leq p} \sum_{i=1}^{n} \mathbb{E}(|R_{i,p}|^3 \mid \mathcal{F}_n) = 8 \max_{1\leq j\leq p} \sum_{i=1}^{n} |\hat{\Theta}_j^\top x_i|^3 \mathbb{E}|\varepsilon_i|^3 \leq 8n\|X\hat{\Theta}\|_\infty^3 \mathbb{E}|\varepsilon_1|^3 \leq B_n.$$

Part (3) follows by

$$\max_{1\leq j\leq p} \sum_{i=1}^{n} \mathbb{E}(|R_{i,j}|^4 \mid \mathcal{F}_n) = 16 \max_{1\leq j\leq p} \sum_{i=1}^{n} |\hat{\Theta}_j^\top x_i|^4 \mathbb{E}|\varepsilon_i|^4 \leq 16n\|X\hat{\Theta}\|_\infty^4 \mathbb{E}|\varepsilon_1|^4 \leq B_n^2.$$

The proof is complete. □

*Proof of Lemma 11.* Recall the event $\mathcal{A}$ defined in Lemma 8, by which we have $\mathbb{P}(\mathcal{A}) \to 1$. Notice that

$$\min_{1\leq j\leq p} n^{-1} \sum_{i=1}^{n} \mathbb{E}(R_{i,j}^2 \mid \mathcal{F}_n) = \min_{1\leq j\leq p} n^{-1} \sum_{i=1}^{n} 4(\hat{\Theta}_j^\top x_i)^2 \sigma_\varepsilon^2 = 4\sigma_\varepsilon^2 \min_{1\leq j\leq p} \|X\hat{\Theta}_j\|_2^2/n. \qquad (D.7)$$

Also notice that, on the event $\mathcal{A}$,

$$\|X_A^\top X_A \hat{\Theta}/m - \mathbb{I}_p\|_\infty \vee \|X_B^\top X_B \hat{\Theta}/m - \mathbb{I}_p\|_\infty \leq \eta.$$

Since $n^{-1}X^\top X - \mathbb{I}_p = (X_A^\top X_A \hat{\Theta}/m - \mathbb{I}_p)/2 + (X_B^\top X_B \hat{\Theta}/m - \mathbb{I}_p)/2$, we have that, on the event $\mathcal{A}$, $\|n^{-1}X^\top X\hat{\Theta} - \mathbb{I}_p\|_\infty \leq \eta$, i.e.,

$$|X_j^\top X\hat{\Theta}_j/n - 1| \leq \eta = o(1) \ \forall 1 \leq j \leq p \ .$$

Thus $|X_j^\top X\hat{\Theta}_j| \geq n(1-\eta) > 0$ for $1 \leq j \leq p$ on $\mathcal{A}$. By Cauchy-Schwarz inequality, $\|X_j\|_2 \|X\hat{\Theta}_j\|_2 \geq |X_j^\top X\hat{\Theta}_j|$ and thus, on the event $\mathcal{A}$,

$$\min_{1 \leq j \leq p} \|X\hat{\Theta}_j\|_2 \geq \frac{n(1-\eta)}{\max_{1 \leq j \leq p} \|X_j\|_2}. \tag{D.8}$$

By the bounded sub-Gaussian norm of $x_{i,j}$, together with Lemma 5.14, Proposition 5.16 in Vershynin (2010) and the union bound,

$$\max_{1 \leq j \leq p} \left| n^{-1} \sum_{i=1}^n (x_{i,j}^2 - \mathbb{E}x_{i,j}^2) \right| = O_P(\sqrt{n^{-1}\log p}).$$

The above display, (D.7) and (D.8) imply that

$$\mathbb{P}\left( \min_{1 \leq j \leq p} n^{-1} \sum_{i=1}^n \mathbb{E}(R_{i,j}^2 \mid \mathcal{F}_n) \geq \frac{(1-\eta)^2 \sigma_\varepsilon^2}{\max_{1 \leq j \leq p} \Sigma_{X,j,j}} \right) \to 1.$$

Since the diagonal entries of $\Sigma_X$ as well as $\sigma_\varepsilon$ are bounded away from zero and infinity, part (1) follows.

Now we show part (2). Let

$$a_i = \mathbf{1}\{i \leq m\} - \mathbf{1}\{i > m\}.$$

Thus, $a_i^2 = 1$. By the bounded sub-Gaussianity of $\varepsilon_i$ and Proposition 5.10 in Vershynin (2010) (applied to the conditional probability measure $\mathbb{P}(\cdot \mid \mathcal{F}_n)$), we have that, on the event $\mathcal{A}$, $\forall t > 0$,

$$\mathbb{P}\left( \left\| n^{-1} \sum_{i=1}^n R_i \right\|_\infty > t \mid \mathcal{F}_n \right) = \mathbb{P}\left( \left\| 2n^{-1} \sum_{i=1}^n a_i \hat{\Theta}_j^\top x_i \varepsilon_i \right\|_\infty > t \mid \mathcal{F}_n \right)$$

$$\leq \sum_{j=1}^p \mathbb{P}\left( \left| \sum_{i=1}^m a_i \hat{\Theta}_j^\top x_i \varepsilon_i \right| > nt \mid \mathcal{F}_n \right)$$

$$\leq p \exp\left( 1 - \frac{K_1 n^2 t^2}{\max_{1 \leq j \leq p} \sum_{i=1}^n (\hat{\Theta}_j^\top x_i)^2} \right) \stackrel{(i)}{=} o_P(1)$$

for some constant $K_1 > 0$, where $(i)$ follows by $\mathbb{P}(\mathcal{A}) \to 1$ and the fact that, on the event $\mathcal{A}$,

$$\max_{1 \leq j \leq p} \sum_{i=1}^{n} (\hat{\Theta}_j^\top x_i)^2 \leq n\|X\hat{\Theta}\|_\infty^2 \leq n\mu^2 = O(n \log(p \vee n)).$$

Since $t$ is arbitrary, part (2) follows.

We apply a similar argument to part (3). Notice that, for $1 \leq j \leq p$,

$$R_{i,j}^2 - \mathbb{E}(R_{i,j}^2 \mid \mathcal{F}_n) = 4(\hat{\Theta}_j^\top x_i)^2(\varepsilon_i^2 - \sigma_\varepsilon^2),$$

where, by Lemma 5.14 in Vershynin (2010), $\varepsilon_i^2 - \sigma_\varepsilon^2$ has a sub-exponential norm upper bounded by some constant $K_2 > 0$. Notice that, on the event $\mathcal{A}$, $\forall t > 0$,

$$\mathbb{P}\left(\max_{1 \leq j \leq p}\left|n^{-1}\sum_{i=1}^{n}[R_{i,j}^2 - \mathbb{E}(R_{i,j}^2 \mid \mathcal{F}_n)]\right| > t \,\Big|\, \mathcal{F}_n\right)$$
$$\leq \sum_{j=1}^{p} \mathbb{P}\left(\left|n^{-1}\sum_{i=1}^{n} 4(\hat{\Theta}_j^\top x_i)^2(\varepsilon_i^2 - \sigma_\varepsilon^2)\right| > t \,\Big|\, \mathcal{F}_n\right)$$
$$\stackrel{(i)}{\leq} 2p\exp\left[-c_0 \min\left(\frac{n^2 t^2}{16K_2^2 \max_{1\leq j\leq p}\sum_{i=1}^{n}(\hat{\Theta}_j^\top x_i)^2}, \frac{nt}{4K_1 \max_{1\leq j\leq p, 1\leq i\leq n}(\hat{\Theta}_j^\top x_i)^2}\right)\right]$$
$$\stackrel{(ii)}{\leq} 2p\exp\left[-c_0 \min\left(\frac{n^2 t^2}{16K_2^2 nO(\log(p \vee n))}, \frac{nt}{4K_1 O(\log(p \vee n))}\right)\right] \stackrel{(iii)}{=} o_P(1)$$

for some universal constant $c_0 > 0$, where $(i)$ follows by Proposition 5.16 in Vershynin (2010) (applied to the conditional probability measure), $(ii)$ follows by the fact that, on $\mathcal{A}$, $\max_{1\leq j\leq p, 1\leq i\leq n}(\hat{\Theta}_j^\top x_i)^2 \leq \mu^2 = O(\log(p \vee n))$ (the constraints in (3.4)) and $(iii)$ holds by $\log^2(p \vee n) = o(n)$. Since $t > 0$ is arbitrary, part (3) follows. The proof is now complete. $\square$

*Proof of Lemma 13.* For notational simplicity, we write $\hat{\Theta}$ instead of $\hat{\Theta}_A$. Notice that, under the assumptions of Theorem 5, the conclusions in Theorem 3.2 of Van de Geer et al. (2014) continue to hold uniformly in $j$:

$$\begin{cases} \max_{1\leq j\leq p} \|\hat{\Theta}_j - \Theta_{*,j}\|_1 = O_P([s_\Theta \vee (s_\beta K_n)]K_n\sqrt{n^{-1}\log p}). \\ \max_{1\leq j\leq p} \left|[\hat{U}_j^\top(\hat{U}_j - \hat{U}_{-j}\hat{\gamma}_j)]/m - \Theta_{*,j,j}\right| = O_P(\sqrt{(K_n^2 s_\beta) \vee s_\Theta} K_n\sqrt{n^{-1}\log p}). \end{cases} \quad (D.9)$$

The proof for (D.9) is largely the same as in Van de Geer et al. (2014) with uniformity in $j$ added in the argument and is thus omitted. The rest of the proof establishes the rates in two steps.

**Step 1: show the rate for $\sup_{t\in[0,1]} \|[\mathbb{I}_p - \hat{\Theta}\hat{H}_A(\hat{\beta}_u - t(\hat{\beta}_u - \beta_*))](\hat{\beta}_u - \beta_*)\|_\infty$.**

By the triangular inequality, we have

$$\sup_{t\in[0,1]} \|[\mathbb{I}_p - \hat{\Theta}\hat{H}_A(\hat{\beta}_u - t(\hat{\beta}_u - \beta_*))](\hat{\beta}_u - \beta_*)\|_\infty$$

$$\leq \underbrace{\|[\mathbb{I}_p - \hat{\Theta}\hat{H}_A(\hat{\beta}_u)](\hat{\beta}_u - \beta_*)\|_\infty}_{J_1} + \underbrace{\sup_{t\in[0,1]} \|\hat{\Theta}[\hat{H}_A(\hat{\beta}_u - t(\hat{\beta}_u - \beta_*)) - \hat{H}_A(\hat{\beta}_u)](\hat{\beta}_u - \beta_*)\|_\infty}_{J_2}. \quad \text{(D.10)}$$

For $j \in \{1, \cdots, p\}$, recall that $\hat{\Theta}_j = (\hat{\Theta}_{j,j}, \hat{\Theta}_{j,-j}^\top)^\top$ with $\hat{\Theta}_{j,-j} = -\hat{\Theta}_{j,j}\hat{\gamma}_j$ and

$$\hat{\Theta}_{j,j} = m/[\hat{U}_j^\top(\hat{U}_j - \hat{U}_{-j}\hat{\gamma}_j)],$$

where $\hat{\gamma}_j = \arg\min_\gamma \|\hat{U}_j - \hat{U}_{-j}\gamma\|_2^2/(2n) + \eta\|\lambda\|_1$. By the KKT condition of this optimization problem, we have $\|\hat{U}_{-j}^\top(\hat{U}_j - \hat{U}_{-j}\hat{\gamma}_j)\|_\infty \leq \eta$. This means that

$$\max_{1\leq j\leq p} \|m^{-1}\hat{U}_{-j}^\top \hat{U}\hat{\Theta}_j\|_\infty = \max_{1\leq j\leq p} |\hat{\Theta}_{j,j}| \cdot \|m^{-1}\hat{U}_{-j}^\top(\hat{U}_j - \hat{U}_{-j}\hat{\gamma}_j)\|_\infty$$

$$\leq \eta \max_{1\leq j\leq p} |\hat{\Theta}_{j,j}|$$

$$= \frac{\eta}{\min_{1\leq j\leq p}[\hat{U}_j^\top(\hat{U}_j - \hat{U}_{-j}\hat{\gamma}_j)]/m}$$

$$\stackrel{(i)}{=} \frac{\eta}{\min_{1\leq j\leq p} \Theta_{*,j,j} + o_P(1)} = O_P(\eta),$$

where $(i)$ holds by (D.9). Also notice that for $1 \leq j \leq p$,

$$m^{-1}\hat{U}_j^\top \hat{U}\hat{\Theta}_j = m^{-1}[\hat{U}_j^\top(\hat{U}_j - \hat{U}_{-j}\hat{\gamma}_j)]\hat{\Theta}_{j,j} = 1$$

by the definition of $\hat{\Theta}_{j,j}$. Hence,

$$\|m^{-1}\hat{U}^\top \hat{U}\hat{\Theta}_j - e_j\|_\infty = O_P(\eta) = O_P(\sqrt{n^{-1}\log p}).$$

Therefore, by Holder's inequality and Lemma 12, we have

$$\|J_1\|_\infty = \|(m^{-1}\hat{U}^\top \hat{U}\hat{\Theta}_j - e_j)\delta\|_\infty \leq \|m^{-1}\hat{U}^\top \hat{U}\hat{\Theta}_j - e_j\|_\infty \|\delta\|_1 = O_P(s_\beta n^{-1}\log p). \quad \text{(D.11)}$$

Notice that $|b'''(z)| = \exp(z)|\exp(z) - 1| \cdot [1 + \exp(z)]^{-3} \leq 1 \; \forall z \in \mathbb{R}$. Hence,

$$\begin{aligned}
\|J_2\|_\infty &= \sup_{t \in [0,1]} \|\hat{\Theta}[\hat{H}_A(\hat{\beta}_u - t(\hat{\beta}_u - \beta_*)) - \hat{H}_A(\hat{\beta}_u)](\hat{\beta}_u - \beta_*)\|_\infty \\
&= \sup_{t \in [0,1]} \max_{1 \leq j \leq p} \left| m^{-1} \sum_{i=1}^m \hat{\Theta}_j^\top x_i x_i^\top (\hat{\beta}_u - \beta_*) \left[ b''(x_i^\top [\hat{\beta}_u - t(\hat{\beta}_u - \beta_*)]) - b''(x_i^\top \hat{\beta}_u) \right] \right| \\
&\leq \sup_{t \in [0,1]} \max_{1 \leq j \leq p} m^{-1} \sum_{i=1}^m \left| \hat{\Theta}_j^\top x_i x_i^\top (\hat{\beta}_u - \beta_*) \right| \cdot \left| b''(x_i^\top [\hat{\beta}_u - t(\hat{\beta}_u - \beta_*)]) - b''(x_i^\top \hat{\beta}_u) \right| \\
&\stackrel{(i)}{\leq} \max_{1 \leq j \leq p} m^{-1} \sum_{i=1}^m \left| \hat{\Theta}_j^\top x_i x_i^\top (\hat{\beta}_u - \beta_*) \right| \cdot \left| x_i^\top (\hat{\beta}_u - \beta_*) \right| \\
&\leq \|X\hat{\Theta}\|_\infty^2 \|X\|_\infty^2 \|\hat{\beta}_u - \beta_*\|_1^2 \stackrel{(ii)}{=} O_P(K_n^4 s_\beta^2 n^{-1} \log p),
\end{aligned} \tag{D.12}$$

where $(i)$ follows by the fact that $\forall u_1, u_2 \in \mathbb{R}$,

$$|b''(u_1) - b''(u_2)| \leq \sup_{u \in \mathbb{R}} |b'''(u)| \cdot |u_1 - u_2| \leq |u_1 - u_2|$$

and $(ii)$ follows by $\|X\|_\infty = O_P(K_n)$, $\|X\Theta_*\|_\infty = O_P(K_n)$ and $\|\hat{\beta}_u - \beta_*\|_1 = O_P(s_\beta \sqrt{n^{-1} \log p})$ (Lemma 12). The rate for $\sup_{t \in [0,1]} \|[\mathbb{I}_p - \hat{\Theta}\hat{H}_A(\hat{\beta}_u - t(\hat{\beta}_u - \beta_*))](\hat{\beta}_u - \beta_*)\|_\infty$ follows by (D.10), together with (D.11) and (D.12).

**Step 2: show the rate for $\|(\hat{\Theta} - \Theta_*)m^{-1} \sum_{i=1}^m s(z_i, \beta_*)\|_\infty$.**

Notice that $|b'(u)| = \exp(u)/[1 + \exp(u)] \leq 1$ and $y_i \in \{0, 1\}$. Hence, $|x_{i,j}[-y_i + b'(x_i^\top \beta_*)]| \leq 2|x_{i,j}|$, which has bounded sub-Gaussian norms. By Proposition 5.10 of Vershynin (2010), we have that $\forall t > 0$,

$$\mathbb{P}\left( \max_{1 \leq j \leq p} \left| \sum_{i=1}^m x_{i,j}[-y_i + b'(x_i^\top \beta_*)] \right| > t \right) \leq \sum_{j=1}^p \mathbb{P}\left( \left| \sum_{i=1}^m x_{i,j}[-y_i + b'(x_i^\top \beta_*)] \right| > t \right) \leq 2p \exp\left[-Ct^2/m\right],$$

where $C > 0$ is a constant depending only on the sub-Gaussian norm of $x_{i,j}$. Hence, $\|\sum_{i=1}^m s(z_i, \beta_*)\|_\infty = O_P(\sqrt{n \log p})$ and by Holder's inequality

$$\left\| (\hat{\Theta} - \Theta_*) m^{-1} \sum_{i=1}^m s(z_i, \beta_*) \right\|_\infty \leq \max_{1 \leq j \leq p} \|\hat{\Theta}_j - \Theta_{*,j}\|_1 \left\| m^{-1} \sum_{i=1}^m s(z_i, \beta_*) \right\|_\infty = O_P\left([s_\Theta \vee (K_n s_\beta)] K_n n^{-1} \log p\right),$$

where the last equality follows by (D.9). $\square$

*Proof of Lemma 14.* The proof is almost the same as that of Lemma 13. We only outline the key steps here. Again, we write $\hat{\Theta}$ instead of $\hat{\Theta}_B$ for notational convenience.

**Step 1: show the rate for $\sup_{t\in[0,1]} \|[\mathbb{I}_p - \hat{\Theta}\hat{H}_B(\hat{\beta}_d - t(\hat{\beta}_d - \beta_*))](\hat{\beta}_d - \beta_*)\|_\infty$.**

Let $K_n = \sqrt{\log(p \vee n)}$. Similar to (D.10), we still have

$$\sup_{t\in[0,1]} \|[\mathbb{I}_p - \hat{\Theta}\hat{H}_B(\hat{\beta}_d - t(\hat{\beta}_d - \beta_*))](\hat{\beta}_d - \beta_*)\|_\infty$$

$$\leq \underbrace{\|[\mathbb{I}_p - \hat{\Theta}\hat{H}_B(\hat{\beta}_u)](\hat{\beta}_d - \beta_*)\|_\infty}_{J_1} + \underbrace{\sup_{t\in[0,1]} \|\hat{\Theta}[\hat{H}_B(\hat{\beta}_d - t(\hat{\beta}_d - \beta_*)) - \hat{H}_B(\hat{\beta}_u)](\hat{\beta}_d - \beta_*)\|_\infty}_{J_2}. \quad \text{(D.13)}$$

Notice that under $H_0$, $\|\hat{\beta}_d - \beta_*\|_1 \leq 2\|\hat{\beta}_u - \beta_*\|_1$ by Lemma 1. Thus, the same argument for (D.11) still yields

$$\|J_1\|_\infty = O_P(s_\beta n^{-1} \log p). \quad \text{(D.14)}$$

Moreover, under $H_0$,

$$\sup_{t\in[0,1]} \|\hat{\beta}_d - \hat{\beta}_u - t(\hat{\beta}_d - \beta_*)\|_1 \leq \|\hat{\beta}_d - \beta_*\|_1 + \|\hat{\beta}_u - \beta_*\|_1 + \sup_{t\in[0,1]} \|t(\hat{\beta}_d - \beta_*)\|_1 \leq 5\|\hat{\beta}_u - \beta_*\|_1. \quad \text{(D.15)}$$

Similar to (D.12), we have that

$$\|J_2\|_\infty = \sup_{t\in[0,1]} \|\hat{\Theta}[\hat{H}_B(\hat{\beta}_d - t(\hat{\beta}_d - \beta_*)) - \hat{H}_B(\hat{\beta}_u)](\hat{\beta}_d - \beta_*)\|_\infty$$

$$\leq \sup_{t\in[0,1]} \max_{1\leq j\leq p} m^{-1} \sum_{i=1}^m \left|\hat{\Theta}_j^\top x_i x_i^\top (\hat{\beta}_d - \beta_*)\right| \cdot \left|x_i^\top (\hat{\beta}_d - \hat{\beta}_u - t(\hat{\beta}_d - \beta_*))\right|$$

$$\overset{(i)}{\leq} \max_{1\leq j\leq p} m^{-1} \sum_{i=1}^m \left|\hat{\Theta}_j^\top x_i x_i^\top (\hat{\beta}_d - \beta_*)\right| \cdot \|X\|_\infty \sup_{t\in[0,1]} \|\hat{\beta}_d - \hat{\beta}_u - t(\hat{\beta}_d - \beta_*)\|_1$$

$$\overset{(ii)}{\leq} \max_{1\leq j\leq p} m^{-1} \sum_{i=1}^m \left|\hat{\Theta}_j^\top x_i x_i^\top (\hat{\beta}_d - \beta_*)\right| \cdot \|X\|_\infty \cdot 5\|\hat{\beta}_u - \beta_*\|_1$$

$$\overset{(iii)}{\leq} 10\|X\hat{\Theta}\|_\infty^2 \|X\|_\infty^2 \|\hat{\beta}_u - \beta_*\|_1^2 = O_P(K_n^4 s_\beta^2 n^{-1} \log p), \quad \text{(D.16)}$$

where $(i)$ follows by Holder's inequality, $(ii)$ follows by (D.15) and $(iii)$ follows by Holder's inequality and $\|\hat{\beta}_d - \beta_*\|_1 \leq 2\|\hat{\beta}_u - \beta_*\|_1$. Hence, we have the same conclusion as Step 1 in the proof of Lemma 13.

**Step 2: show the rate for $\|(\hat{\Theta} - \Theta_*)m^{-1}\sum_{i=1}^m s(z_i, \beta_*)\|_\infty$.**

Using the same argument as in Step 2 of the proof of Lemma 13, we have

$$\left\|(\hat{\Theta} - \Theta_*)m^{-1}\sum_{i=m+1}^{n} s(z_i, \beta_*)\right\|_\infty = O_P\left([s_\Theta \vee (K_n s_\beta)]K_n n^{-1}\log p\right).$$

The proof is complete. □

*Proof of Lemma 15.* Notice that $\|X\|_\infty = O_P(K_n)$ and $\|X\Theta_*\|_\infty = O_P(K_n)$. Observe that

$$\max_{1\leq j\leq p} m^{-1}\sum_{i=1}^{m}(\hat{R}_{i,j} - R_{i,j})^2 \tag{D.17}$$

$$= \max_{1\leq j\leq p} 4m^{-1}\sum_{i=1}^{m}\left(\hat{\Theta}_{A,j}^\top x_i[-y_i + b'(x_i^\top \hat{\beta}_u)] - \Theta_{*,j}^\top x_i[-y_i + b'(x_i^\top \beta_*)]\right)^2$$

$$\stackrel{(i)}{\leq} 8\max_{1\leq j\leq p} m^{-1}\sum_{i=1}^{m}\left[(\hat{\Theta}_{A,j} - \Theta_{*,j})^\top x_i\right]^2 \left(-y_i + b'(x_i^\top \hat{\beta}_u)\right)^2 \tag{D.18}$$

$$+ 8\max_{1\leq j\leq p} m^{-1}\sum_{i=1}^{m}(\Theta_{*,j}^\top x_i)^2 \left(b'(x_i^\top \hat{\beta}_u) - b'(x_i^\top \beta_*)\right)^2$$

$$\stackrel{(ii)}{\leq} 8\|X\|_\infty^2 \max_{1\leq j\leq p}\|\hat{\Theta}_{A,j} - \Theta_{*,j}\|_1^2 m^{-1}\sum_{i=1}^{m}\left(-y_i + b'(x_i^\top \hat{\beta}_u)\right)^2 \tag{D.19}$$

$$+ 8\|X\Theta_*\|_\infty^2 m^{-1}\sum_{i=1}^{m}\left(b'(x_i^\top \hat{\beta}_u) - b'(x_i^\top \beta_*)\right)^2$$

$$\stackrel{(iii)}{=} O_P([s_\Theta^2 \vee (K_n^2 s_\beta^2)]K_n^4 n^{-1}\log p) + O_P(K_n^2)m^{-1}\sum_{i=1}^{m}\left(b'(x_i^\top \hat{\beta}_u) - b'(x_i^\top \beta_*)\right)^2,$$

where $(i)$ holds by the elementary inequality $(a+b)^2 \leq 2a^2 + 2b^2$, $(ii)$ follows by Holder's inequality and $(iii)$ follows by (D.9) in the proof of Lemma 13, $\|X\|_\infty = O_P(K_n)$, $\|X\Theta_*\|_\infty = O_P(K_n)$ and the fact that $|y_i| \leq 1$ and $\sup_{u\in\mathbb{R}}|b'(u)| \leq 1$.

Let $\delta = \hat{\beta}_u - \beta_*$ and notice that, by Lemma 12, $\|\delta\|_1 = O_P(s_\beta\sqrt{n^{-1}\log p})$. Notice that $\forall u \in \mathbb{R}$, $|b''(u)| = \exp(u)/[1+\exp(u)]^2 \leq 1$. We have

$$|b'(x_i^\top \hat{\beta}_u) - b'(x_i^\top \beta_*)| \leq |x_i'\delta|$$

and thus

$$m^{-1}\sum_{i=1}^{m}\left(b'(x_i^\top \hat{\beta}_u) - b'(x_i^\top \beta_*)\right)^2 \leq m^{-1}\sum_{i=1}^{m}(x_i^\top \delta)^2 \stackrel{(i)}{\leq} \|X_A\|_\infty^2\|\delta\|_1^2 = O_P(K_n^2 s_\beta^2 n^{-1}\log p), \tag{D.20}$$

where $(i)$ follows by Holder's inequality. Thus, (D.17) and (D.20) imply

$$\max_{1\leq j\leq p} m^{-1} \sum_{i=1}^{m}(\hat{R}_{i,j} - R_{i,j})^2 = O_P([s_\Theta^2 \vee (K_n^2 s_\beta^2)]K_n^4 n^{-1}\log p).$$

A similar argument yields $\max_{1\leq j\leq p} m^{-1} \sum_{i=m+1}^{n}(\hat{R}_{i,j} - R_{i,j})^2 = O_P([s_\Theta^2 \vee (K_n^2 s_\beta^2)]K_n^4 n^{-1}\log p)$. We conclude the proof by observing

$$\max_{1\leq j\leq p} n^{-1}\sum_{i=1}^{n}(\hat{R}_{i,j} - R_{i,j})^2 \leq \frac{1}{2}\max_{1\leq j\leq p} m^{-1}\sum_{i=1}^{m}(\hat{R}_{i,j} - R_{i,j})^2 + \frac{1}{2}\max_{1\leq j\leq p} m^{-1}\sum_{i=m+1}^{n}(\hat{R}_{i,j} - R_{i,j})^2.$$

□

# E Technical tools

**Lemma 16.** *Let $X$ and $Y$ be two random vectors. Then $\forall t, \varepsilon > 0$, $|\mathbb{P}(\|X\|_\infty > t) - \mathbb{P}(\|Y\|_\infty > t)| \leq \mathbb{P}(\|X - Y\|_\infty > \varepsilon) + \mathbb{P}(\|Y\|_\infty \in (t-\varepsilon, t+\varepsilon])$.*

*Proof of Lemma 16.* The result holds by the following observations using the triangular inequality: (1) $\mathbb{P}(\|X\|_\infty > t) \leq \mathbb{P}(\|X-Y\|_\infty > \varepsilon) + \mathbb{P}(\|Y\|_\infty > t-\varepsilon) = \mathbb{P}(\|X-Y\|_\infty > \varepsilon) + \mathbb{P}(\|Y\|_\infty > t) + \mathbb{P}(\|Y\|_\infty \in (t-\varepsilon, t])$ and (2) $\mathbb{P}(\|X\|_\infty > t) \geq \mathbb{P}(\|Y\|_\infty > t+\varepsilon) - \mathbb{P}(\|X-Y\|_\infty > \varepsilon) = \mathbb{P}(\|Y\|_\infty > t) - \mathbb{P}(\|Y\|_\infty \in (t, t+\varepsilon]) - \mathbb{P}(\|X-Y\|_\infty > \varepsilon)$. □

**Lemma 17.** *Let $X$ and $Y$ be two random vectors and $\mathcal{F}$ and $\mathcal{G}$ two $\sigma$-algebras. Define $F_X(x) = \mathbb{P}(\|X\|_\infty \leq x \mid \mathcal{F})$ and $F_Y(x) = \mathbb{P}(\|Y\|_\infty \leq x \mid \mathcal{G})$. Then $\forall \varepsilon > 0$, $\sup_{\alpha \in (0,1)} |\mathbb{P}(\|X\|_\infty > F_Y^{-1}(1-\alpha) \mid \mathcal{F}) - \alpha| \leq \varepsilon + \mathbb{P}(\sup_{x\in\mathbb{R}} |F_X(x) - F_Y(x)| > \varepsilon \mid \mathcal{F})$.*

*Proof of Lemma 17.* For simplicity, we use $\mathbb{P}_\mathcal{F}(\cdot)$ to denote $\mathbb{P}(\cdot \mid \mathcal{F}_n)$. Fix $\alpha \in (0,1)$ and notice that

$$\mathbb{P}_\mathcal{F}\left(\|X\|_\infty > F_Y^{-1}(1-\alpha)\right) \tag{E.1}$$

$$\leq \mathbb{P}_\mathcal{F}\left(\|X\|_\infty > F_Y^{-1}(1-\alpha) \text{ and } \sup_{x\in\mathbb{R}}|F_X(x) - F_Y(x)| \leq \varepsilon\right) + \mathbb{P}_\mathcal{F}\left(\sup_{x\in\mathbb{R}}|F_X(x) - F_Y(x)| > \varepsilon\right) \tag{E.2}$$

$$\overset{(i)}{\leq} \mathbb{P}_\mathcal{F}\left(\|X\|_\infty > F_X^{-1}(1-\alpha-\varepsilon)\right) + \mathbb{P}_\mathcal{F}\left(\sup_{x\in\mathbb{R}}|F_X(x) - F_Y(x)| > \varepsilon\right) \tag{E.3}$$

$$= \alpha + \varepsilon + \mathbb{P}_\mathcal{F}\left(\sup_{x\in\mathbb{R}}|F_X(x) - F_Y(x)| > \varepsilon\right), \tag{E.4}$$

where $(i)$ follows from Lemma A.1(ii) in Romano and Shaikh (2012) (if $\sup_{x\in\mathbb{R}}[F_Y(x) - F_X(x)] \leq \varepsilon$ then $F_X^{-1}(1-\alpha-\varepsilon) \leq F_Y^{-1}(1-\alpha)$). Also notice that

$$
\begin{aligned}
\mathbb{P}_{\mathcal{F}}\left(\|X\|_\infty > F_Y^{-1}(1-\alpha)\right) &\geq \mathbb{P}_{\mathcal{F}}\left(\|X\|_\infty > F_Y^{-1}(1-\alpha) \text{ and } \sup_{x\in\mathbb{R}}|F_X(x) - F_Y(x)| \leq \varepsilon\right) \\
&\overset{(i)}{\geq} \mathbb{P}_{\mathcal{F}}\left(\|X\|_\infty > F_X^{-1}(1-\alpha+\varepsilon) \text{ and } \sup_{x\in\mathbb{R}}|F_X(x) - F_Y(x)| \leq \varepsilon\right) \\
&\geq \mathbb{P}_{\mathcal{F}}\left(\|X\|_\infty > F_X^{-1}(1-\alpha+\varepsilon)\right) - \mathbb{P}_{\mathcal{F}}\left(\sup_{x\in\mathbb{R}}|F_X(x) - F_Y(x)| > \varepsilon\right) \\
&= \alpha - \varepsilon - \mathbb{P}_{\mathcal{F}}\left(\sup_{x\in\mathbb{R}}|F_X(x) - F_Y(x)| > \varepsilon\right)
\end{aligned}
\quad (\text{E.5})
$$

where $(i)$ follows from Lemma A.1(ii) in Romano and Shaikh (2012) (if $\sup_{x\in\mathbb{R}}[F_X(x) - F_Y(x)] \leq \varepsilon$ then $F_Y^{-1}(1-\alpha) \leq F_X^{-1}(1-\alpha+\varepsilon)$). The desired result follows by (E.4) and (E.5). □

**Lemma 18.** *Let $Y = (Y_1, \cdots, Y_p)^\top$ be a random vector and $\mathcal{F}$ a $\sigma$-algebra. If $\mathbb{E}(Y \mid \mathcal{F}) = 0$, $Y \mid \mathcal{F}$ is Gaussian and $\min \mathbb{E}(Y_j^2 \mid \mathcal{F}) \geq b$ a.s. for some constant $b > 0$, then there exists a constant $C_b > 0$ depending only on $b$ such that $\forall \varepsilon > 0$.*

$$
\sup_{x\in\mathbb{R}} \mathbb{P}\left(\|Y\|_\infty \in (x-\varepsilon, x+\varepsilon] \mid \mathcal{F}\right) \leq C_b \varepsilon \sqrt{\log p} \quad a.s.
$$

*Proof of Lemma 18.* By Nazarov's anti-concentration inequality (Lemma A.1 in Chernozhukov et al. (2014)), there exists a constant $C_b^\top$ depending only on $b$ such that almost surely, $\sup_{x\in\mathbb{R}} \mathbb{P}(\max_{1\leq j\leq p} Y_j \in (x-\varepsilon, x+\varepsilon] \mid \mathcal{F}) \leq 2C_b^\top \varepsilon\sqrt{\log p}$ and $\sup_{x\in\mathbb{R}} \mathbb{P}(\max_{1\leq j\leq p}(-Y_j) \in (x-\varepsilon, x+\varepsilon] \mid \mathcal{F}) \leq 2C_b^\top \varepsilon\sqrt{\log p}$.

Since $\|Y\|_\infty = \max\{\max_{1\leq j\leq p} Y_j, \max_{1\leq j\leq p}(-Y_j)\}$, the desired result follows by $\sup_{x\in\mathbb{R}} \mathbb{P}(\|Y\|_\infty \in (x-\varepsilon, x+\varepsilon] \mid \mathcal{F}) \leq \sup_{x\in\mathbb{R}} \mathbb{P}(\max_{1\leq j\leq p} Y_j \in (x-\varepsilon, x+\varepsilon] \mid \mathcal{F}) + \sup_{x\in\mathbb{R}} \mathbb{P}(\max_{1\leq j\leq p}(-Y_j) \in (x-\varepsilon, x+\varepsilon] \mid \mathcal{F}) \leq 4C_b^\top \varepsilon\sqrt{\log p}$. □

**Lemma 19.** *Let $X$ and $Y$ be two sub-Gaussian random variables. Then $XY$ is sub-exponential.*

*Proof of Lemma 19.* Since $X$ and $Y$ are sub-Gaussian, there exist constants $C_1, C_2 > 0$ such that $\mathbb{P}(|X| > z) \leq \exp(-C_1 z^2)$ and $\mathbb{P}(|Y| > z) \leq \exp(-C_2 z^2)$ $\forall z > 0$.

Let $C = \min\{C_1, C_2\}$. Fix any constant $t > 0$. We have that

$$
\begin{aligned}
\mathbb{P}(|XY| > t) &\leq \mathbb{P}\left(|XY| > t \text{ and } |Y| > \sqrt{t}\right) + \mathbb{P}\left(|XY| > t \text{ and } |Y| \leq \sqrt{t}\right) \\
&\leq \mathbb{P}\left(|Y| > \sqrt{t}\right) + \mathbb{P}\left(|X| > \sqrt{t}\right) \leq \exp(-C_2 t) + \exp(-C_1 t) \leq 2\exp(-Ct).
\end{aligned}
$$

Thus, $XY$ is sub-exponential. □